\documentstyle[prd,aps]{revtex}
\begin{document}
\begin{flushright}
OCU-PHYS 177 \\ 
July 2000 
\end{flushright}
\begin{center} 
{\Large Out of equilibrium O (N) linear-sigma system --- 
Construction of perturbation \\ theory with gap- and 
Boltzmann-equations} 
\end{center} 

\hspace*{2ex} 

\hspace*{2ex}

\begin{center} 
{\large A. Ni\'{e}gawa}\footnote{Electronic address: 
niegawa@sci.osaka-cu.ac.jp} \\ 
Department of Physics, Osaka City University, 
Sumiyoshi-ku, Osaka 558-8585, JAPAN 
\end{center} 

\hspace*{2ex}

\begin{center} 
{\large Abstract} 
\end{center} 
We establish from first principles a perturbative framework that 
allows us to compute reaction rates for processes taking place in 
nonequilibrium $O (N)$ linear-sigma systems in broken phase. The 
system of our concern is quasiuniform system near equilibrium or 
nonequilibrium quasistationary system. We employ the 
closed-time-path formalism and use the so-called gradient 
approximation. No further approximation is introduced. 
In the course of construction of the framework, we obtain the gap 
equation that determines the effective masses of $\pi$ and of 
$\sigma$, and the generalized Boltzmann equation that describes the 
evolution of the number-density functions of $\pi$ and of $\sigma$. 
\hspace*{2ex} 
\begin{flushleft}
11.10.Wx, 11.30.Qc, 11.30.Rd 
\end{flushleft}
\widetext 
\section{Introduction} 
\setcounter{equation}{0}
\setcounter{section}{1} 
\def\theequation{\mbox{\arabic{section}.\arabic{equation}}} 
Lattice Quantum Chromodynamics (QCD) results indicate that chiral 
symmetry is spontaneously broken at $T \sim 150$ MeV. Such 
temperatures may be reached in relativistic heavy-ion collisions. 
Thus, the time evolution of the chiral phase transition may be 
traced through observing predicted consequences, e.g., in the 
production rates of dilepton and of photon, etc. \cite{re-b}, or the 
formation of disoriented-chiral-condensate (DCC) domains (see, 
e.g., \cite{bjo}). 

In this paper, for an effective low-energy model for QCD, we take 
the $O (N)$ linear-sigma model. The case of $N = 4$ is of practical 
interest. The ultimate goal is to construct a framework for 
analyzing how the phase transition proceeds through DCC's. A 
numerous work was reported so far on the \lq\lq static'' properties 
of this model or its variants. (See, e.g., \cite{pis,chiku}. Earlier 
work is quoted therein.) Since Rajagopal and Wilczek \cite{51} 
proposed that heavy-ion collisions can generate large DCC domains, 
analyses of dynamical evolution of the system from a symmetric phase 
to a broken phase have energized. The analysis in \cite{51} has been 
refined in \cite{52}, and has been extended \cite{53} by 
incorporating more realistic descriptions of heavy-ion dynamics. The 
effects of quasiparticle excitations are studied in \cite{54}. 
Refinements employing realistic initial conditions are made, e.g., 
in \cite{55}. Since then, much work has been devoted to the analysis 
by systematically taking quantum and medium effects into account: 
For example, analysis on the basis of density-matrix formalism is 
made in \cite{56}, large-$N$ limit of the $O (N)$ linear-sigma model 
has been studied in \cite{57}, closed-time-path (CTP) formalism of 
nonequilibrium dynamics is employed in \cite{58}, Calderira-Leggett 
theory is applied in \cite{59}, a self-consistent-variational 
approach has been taken in \cite{60}, the time evolution of a 
particle distribution is studied in \cite{61}, and relaxation rate 
for long-wavelength fluctuations are analyzed in \cite{62}. (For 
other related works, see, e.g., \cite{63}.) Different assumptions 
and approximations are employed in these analyses. 

In this paper, as a first step toward the goal, we lay down {\em 
from first principles} a perturbative framework on the basis of a 
loop-expansion scheme. Only approximation we use is the so-called 
gradient approximation (see below). We use the standard framework of 
nonequilibrium statistical quantum-field theory that is formulated 
by employing the closed-time path, $- \infty \to + \infty \to - 
\infty$, in a complex-time plane \cite{sch,chou,lan} , which is 
referred to as the CTP formalism. 

Throughout this paper, we are interested in quasiuniform systems 
near equilibrium or nonequilibrium quasistationary systems. Such 
systems are characterized by two different spacetime scales; 
microscopic or quantum-field-theoretical and macroscopic or 
statistical. The first scale, the microscopic correlation scale, 
characterizes the reaction taking place in the system, while the 
second scale measures the relaxation of the system. For a weak 
coupling theory, in which we are interested in this paper, the 
former scale is much smaller than the latter scale.\footnote{
It should be noted, however, that, as the system approaches the 
critical point of the phase transition, the microscopic correlation 
scale diverges. Thus, the formalism developed in this paper applies 
to the systems away from the critical point.} In a derivative 
expansion with respect to macroscopic spacetime coordinates $X$, we 
use the gradient approximation throughout: 
\begin{equation} 
F (X, ...) \simeq F (Y, ...) + (X - Y)^\mu \partial_{Y^\mu} F (Y, 
...) \, . 
\label{gra} 
\end{equation} 
Let $\Delta (x, y)$ be a generic propagator. For the system of our 
concern, $\Delta (x, y)$, with $x - y$ fixed, does not change 
appreciably in $(x + y) / 2$. We refer the first term on the R.H.S. 
to as the leading part and the second term to as the gradient part. 
The self-energy part $\Sigma (x, y)$ enjoys a similar property. 
Thus, we choose $x - y$ as the microscopic coordinates while $X 
\equiv (x + y) / 2$ as the macroscopic coordinates. 

The plan of the paper is as follows. The $O (N)$ linear-sigma model 
is introduced in Sec.~II and the forms of retarded and advanced bare 
propagators are given in Sec.~III. In Sec.~IV, a perturbative 
framework is constructed from first principles. The framework thus 
constructed allows us to compute reaction rates by using the 
reaction-rate formula \cite{oka}. In Sec.~V, a quasiparticle 
representation of the propagator is given. In Sec.~VI, after 
constructing the self-energy-part resummed propagator, the gap 
equation and the generalized Boltzmann equation are derived. Section 
VII is devoted to conclusion and outlook. Concrete derivation of 
various formula used in the text is made in Appendices. 
\section{$O (N)$ linear-sigma model} 
\setcounter{equation}{0}
\setcounter{section}{2} 
\def\theequation{\mbox{\arabic{section}.\arabic{equation}}} 
The Lagrangian (density) of the $O (N)$ linear-sigma model reads 
\begin{equation} 
{\cal L} = \frac{1}{2} \left[ (\partial \vec{\phi}_B )^2 - m^2_B 
\vec{\phi}^{\, 2}_B \right]  - \frac{\lambda_B}{4 !} ( 
\vec{\phi}^{\, 2}_B)^2 + \displaystyle{ 
\raisebox{0.9ex}{\scriptsize{$t$}}} \mbox{\hspace{-0.1ex}} 
\vec{H}_B \cdot \vec{\phi}_B \, , 
\label{on-lag} 
\end{equation} 
where $\vec{\phi}_B = (\phi_B^1, \phi_B^2, ... , \phi_B^N)$. When 
$m^2_B < 0$, ${\cal L}$ describes the system whose ground state is 
in a broken phase in the classical limit. ${\vec H}_B$ is an 
external field, which explicitly breaks $O (N)$ symmetry. Noticing 
the fact that a renormalization scheme for the symmetric phase 
($m_B^2 > 0$) works \cite{lee} as it is for the broken phase ($m_B^2 
< 0$), we introduce renormalized quantities, $\vec{\phi}_B = 
\sqrt{Z} \vec{\phi}$, $m_B^2 = Z_m m^2$, and $\lambda_B = Z_\lambda 
\lambda$, in terms of which ${\cal L}$ reads 
\begin{eqnarray} 
{\cal L} (\vec{\phi}) & = & \frac{1}{2} \left[ (\partial \vec{\phi} 
)^2 - m^2 \vec{\phi}^{\, 2} \right] - \frac{\lambda}{4 !} ( 
\vec{\phi}^{\, 2})^2 + \displaystyle{ 
\raisebox{0.9ex}{\scriptsize{$t$}}} \mbox{\hspace{-0.1ex}} \vec{H} 
\cdot \vec{\phi} + \frac{1}{2} (Z - 1) (\partial \vec{\phi} 
)^2 - \frac{Z_m Z - 1}{1} m^2 \vec{\phi}^{\, 2} \nonumber \\ 
& & - (Z_\lambda Z^2 - 1) \frac{\lambda}{4 !} ( \vec{\phi}^{\, 2})^2 
\, , 
\label{ki-ki} 
\end{eqnarray} 
where $\vec{H} = \sqrt{Z} \vec{H}_B$. 

The system in the broken phase is governed by the \lq\lq sifted 
Lagrangian:'' 
\[ 
{\cal L} (\vec{\phi} (x); \vec{\varphi} (x)) \equiv {\cal L} 
(\vec{\phi} (x) + \vec{\varphi} (x)) - \frac{\partial {\cal L} 
(\vec{\varphi} (x))}{\partial \vec{\varphi} (x)} \cdot  \vec{\phi} 
(x) \, . 
\] 
To avoid too many notations, for ${\cal L} (\vec{\phi} (x); 
\vec{\varphi} (x))$, we have used the same letter \lq\lq ${\cal 
L}$'' as in Eqs.~(\ref{on-lag}) and (\ref{ki-ki}). $\vec{\varphi} 
(x)$ is the (classical) condensate or order-parameter fields and 
$\vec{\phi} (x)$ is the quantum fields, which describes the 
fluctuation around $\vec{\varphi} (x)$. Straightforward manipulation 
yields 
\begin{eqnarray} 
{\cal L} (\vec{\phi} (x); \vec{\varphi} (x)) & = & {\cal L}_q' + 
{\cal L}_{r c} + ... \, , 
\label{2.3} 
\\ 
{\cal L}_q' & = & \frac{1}{2} \left[ ( \partial \vec{\phi})^2 - 
m^2 \vec{\phi}^{\, 2}  \right] - \frac{\lambda}{12} \left[ 
\vec{\varphi}^2 \vec{\phi}^{\, 2} + 2 ( 
\displaystyle{ \raisebox{0.9ex}{\scriptsize{$t$}}} 
\mbox{\hspace{-0.1ex}} \vec{\varphi} \cdot \vec{\phi})^2 \right] - 
\frac{\lambda}{3 !} ( \displaystyle{ 
\raisebox{0.9ex}{\scriptsize{$t$}}} \mbox{\hspace{-0.1ex}} 
\vec{\varphi} \cdot \vec{\phi}) \, \vec{\phi}^{\, 2} - 
\frac{\lambda}{4 !} ( \vec{\phi}^{\, 2})^2 \, , \nonumber \\ 
{\cal L}_{r c} & = & \frac{1}{2} (Z - 1) ( \partial \vec{\phi})^2 
- \frac{1}{2} (Z_m Z - 1) m^2 \vec{\phi}^{\, 2} \nonumber \\ 
&& - (Z_\lambda Z^2 
- 1) \left[ \frac{\lambda}{12} \left\{ \vec{\varphi}^{\, 2} 
\vec{\phi}^{\, 2} + 2 ( \displaystyle{ 
\raisebox{0.9ex}{\scriptsize{$t$}}} \mbox{\hspace{-0.1ex}} 
\vec{\varphi} \cdot \vec{\phi})^2 \right\} + \frac{\lambda}{3 !} 
( \displaystyle{ \raisebox{0.9ex}{\scriptsize{$t$}}} 
\mbox{\hspace{-0.1ex}} \vec{\varphi} \cdot \vec{\phi}) \, 
\vec{\phi}^{\, 2} + \frac{\lambda}{4 !} ( \vec{\phi}^{\, 2})^2 
\right] \, . 
\label{L-q} 
\end{eqnarray} 
In Eq.~(\ref{2.3}), \lq $...$' stands for the terms that includes 
only c-number field $\vec{\varphi}$, which is not necessary for the 
present purpose, but plays a role when spacetime evolution of the 
system is studied. We ignore \lq $...$' throughout in the sequel. It 
is to be noted that ${\cal L}$ enjoys $O (N - 1)$ symmetry. 

For obtaining an efficient or rather physically-sensible 
(perturbative) scheme, we introduce \cite{chiku,wein,fen,law} weakly 
$x$-dependent masses, $M_\pi(x)$ and $M_\sigma (x)$, where \lq $x$' 
is macroscopic coordinates. A consistent perturbative scheme is 
obtained by assuming that, when compared to $m^2$, $\chi_\xi (x) 
\equiv M_\xi^2 (x) - m^2$ $(\xi = \pi, \sigma)$ are one-order higher 
in the loop expansion. How to determine $M_\pi^2 (x)$ and 
$M_\sigma^2 (x)$ will be discussed in Sec.~VIB. We then rewrite 
${\cal L} (\vec{\phi} (x); \vec{\varphi} (x))$ in the form, 
\begin{equation} 
{\cal L} (\vec{\phi} (x); \vec{\varphi} (x)) = {\cal L}_0 
(\vec{\phi} (x); \vec{\varphi} (x)) + {\cal L}_{int} + {\cal L}_{r 
c} + {\cal L}_{m c} \, . 
\label{owari} 
\end{equation} 
Here ${\cal L}_{r c}$ is as in Eq.~(\ref{L-q}) and 
\begin{eqnarray} 
{\cal L}_0 (\vec{\phi} (x); \vec{\varphi} (x)) & = & \frac{1}{2} 
\int d^{\, 4} y \, \displaystyle{ 
\raisebox{0.9ex}{\scriptsize{$t$}}} \mbox{\hspace{-0.1ex}} 
\vec{\phi} (x) \, {\bf \Delta}^{- 1} (x, y) \, \vec{\phi} (y) \, , 
\label{2.8} \\ 
{\cal L}_{int} & = & - \frac{\lambda}{3 !} ( \displaystyle{ 
\raisebox{0.9ex}{\scriptsize{$t$}}} \mbox{\hspace{-0.1ex}} 
\vec{\varphi} \cdot \vec{\phi}) \vec{\phi}^{\, 2} - \frac{\lambda}{4 
!} ( \vec{\phi}^{\, 2})^2 \equiv {\cal L}_{int}^{(3)} + {\cal 
L}_{int}^{(4)} \, , 
\label{2.6d} \\ 
{\cal L}_{m c} & = & \frac{1}{2} \displaystyle{ 
\raisebox{0.9ex}{\scriptsize{$t$}}} \mbox{\hspace{-0.1ex}} 
\vec{\phi} (x) \left[ \chi_\pi (x) {\bf P}_\pi (x) + \chi_\sigma (x) 
{\bf P}_\sigma (x) \right] \vec{\phi} (x) \, , 
\label{teigi-yo} 
\end{eqnarray} 
where 
\begin{eqnarray} 
{\bf \Delta}^{- 1} (x, y) & \equiv & - \left[ \left( \partial^2 + 
{\cal M}_\pi^2 (x) \right) {\bf P}_\pi (x) + \left( \partial^2 + 
{\cal M}_\sigma^2 (x) \right) {\bf P}_\sigma (x) \right] \delta^{\, 
4} (x - y) \, , 
\label{inv} \\ 
{\bf P}_\pi (x) & = & {\bf I} - | \hat{\varphi} (x) \rangle \langle 
\hat{\varphi} (x) | \, , \;\;\;\;\;\; {\bf P}_\sigma (x) = | 
\hat{\varphi} (x) \rangle \langle \hat{\varphi} (x) | \, , 
\label{2.8d} \\ 
{\cal M}_\pi^2 (X) & \equiv & M_\pi^2 (X) + \frac{\lambda}{6} 
\vec{\varphi}^{\, 2} (X) \, , \;\;\;\; {\cal M}_\sigma^2 (X) \equiv 
M_\sigma^2 (X) + \frac{\lambda}{2} \vec{\varphi}^{\, 2} (X) \, . 
\label{mass} 
\end{eqnarray} 

\noindent 
In these equations, the boldface letters denote $N \times N$ 
matrices act on the real vector space, ${\bf P}_\pi$ (${\bf 
P}_\sigma$) is the projection operator onto the $\pi$- 
$(\sigma)$-subspace, and ${\bf I}$ is an unit matrix. $| 
\hat{\varphi} (x) \rangle$ is a unit 
vector along $\vec{\varphi} (x)$, and $\langle \hat{\varphi} (x)|$ 
is an adjoint of $| \hat{\varphi} (x) \rangle$. 

The construction of perturbation theory based on ${\cal L}$, 
Eq.~(\ref{owari}), starts with constructing the Fock space of the 
quanta described by $\vec{\phi}$, which is defined \lq\lq on 
$\vec{\varphi} (x)$.'' As stated in Sec.~I, we are concerned about 
the systems which are not far from equilibrium states or from 
stationary states. Then, the theory to be developed may be applied 
to the case where $\vec{\varphi} (x)$ changes slowly. In one word, 
\lq $x$' of $\vec{\varphi} (x)$ is macroscopic coordinates. 
\section{Retarded and advanced propagators} 
\setcounter{equation}{0}
\setcounter{section}{3} 
\def\theequation{\mbox{\arabic{section}.\arabic{equation}}} 
For the purpose of later use, we construct retarded and 
advanced propagators. The retarded (advanced) propagator ${\bf 
\Delta}_R$ $({\bf \Delta}_A)$ is an inverse of ${\bf \Delta}^{- 1} 
(x, y)$, Eq.~(\ref{inv}), under the retarded (advanced) 
boundary condition. Derivation of the form of ${\bf \Delta}_R$ and 
${\bf \Delta}_A$ in the gradient approximation is straightforward 
and, here, we display the result: 
\begin{eqnarray} 
{\bf \Delta}_{R (A)} (x, y) & \simeq & {\bf P}_\xi (x) {\Delta}_{R 
(A)}^{(\xi)} (x, y) {\bf P}_\xi (y) + {\bf \Delta}^{(1)}_{R (A)} 
(x, y) \, , 
\label{RA-de} \\ 
{\Delta}_{R (A)}^{(\xi)} (x, y) & = & \int \frac{d^{\, 4} P}{(2 
\pi)^4} \, e^{- i P \cdot (x - y)} {\Delta}^{(\xi)}_{R (A)} (X; 
P) \, , 
\label{3.1} \\ 
{\bf \Delta}^{(1)}_{R (A)} (x, y) & = & 2 i \left( | \varphi (X) 
\rangle 
\stackrel{\leftrightarrow}{\partial}_{X_\mu} \langle \varphi (X) | 
\right) \int \frac{d^{\, 4} P}{(2 \pi)^4} \, e^{- i P \cdot (x - y)} 
P_\mu \Delta_R^{(\pi)} (X; P) \Delta_R^{(\sigma)} (X; P) \, . 
\label{ana} 
\end{eqnarray} 
In Eq.~(\ref{RA-de}), summation over $\xi$ runs over $\pi$ and 
$\sigma$, and, in Eqs.~(\ref{3.1}) and (\ref{ana}), $X = (x + y) / 
2$, $\stackrel{\leftrightarrow}{\partial}_X \equiv 
\stackrel{\rightarrow}{\partial}_X - 
\stackrel{\leftarrow}{\partial}_X$, and 
\[ 
\Delta_{R (A)}^{(\xi)} (X; P) \equiv \frac{1}{P^2 - {\cal 
M}^2_\xi (X) \pm i \epsilon(p_0) 0^+} \;\;\;\;\;\; (\xi = \pi, 
\sigma) \, . 
\] 

That the gradient part ${\bf \Delta}^{(1)}_{R (A)} (x, y)$ appears 
is a reflection of the fact that the internal reference frame, which 
defines $\sigma$ mode and three-$\pi$ modes, at the spacetime point 
$x$ is different from that at the point $y$. Let us see the meaning 
of $| \hat{\varphi} \rangle P \cdot 
\stackrel{\leftrightarrow}{\partial} \langle \hat{\varphi}|$ in 
${\bf \Delta}^{(1)}_{R (A)}$. From Eq.~(\ref{2.8d}), we can easily 
see that 
\begin{eqnarray*} 
{\bf P}_\sigma |\hat{\varphi} \rangle P \cdot \partial 
\langle \hat{\varphi}| {\bf P}_\pi & = & 
|\hat{\varphi} \rangle P \cdot \partial 
\langle \hat{\varphi}| \, , \\ 
{\bf P}_\xi |\hat{\varphi} \rangle P \cdot \partial 
\langle \hat{\varphi}| {\bf P}_\zeta & = & 0 
\;\;\;\;\; ((\xi, \zeta) \neq (\sigma, \pi)) \, ,  
\end{eqnarray*} 
where use has been made of $\hat{\varphi} \cdot \partial 
\hat{\varphi} = 0$ $(\hat{\varphi} \equiv \vec{\varphi} / 
|\vec{\varphi}|)$. $| \hat{\varphi} \rangle P \cdot 
\stackrel{\leftarrow}{\partial} \langle \hat{\varphi}|$ enjoys 
similar property. Then, $|\hat{\varphi} \rangle P \cdot \partial 
\langle \hat{\varphi}|$ and $| \hat{\varphi} \rangle P \cdot 
\stackrel{\leftarrow}{\partial} \langle \hat{\varphi}|$ \lq\lq 
induce'' transition between $\pi$ mode and $\sigma$ mode. 
\section{Construction of perturbative framework} 
\setcounter{equation}{0}
\setcounter{section}{4} 
\def\theequation{\mbox{\arabic{section}.\arabic{equation}}} 
\subsection{Preliminary} 
The CTP formalism is formulated \cite{chou} by introducing an 
oriented closed-time path $C$ $(= C_1 + C_2)$ in a complex-time 
plane, that goes from $- \infty$ to $+ \infty$ $(C_1)$ and then 
returns from $+ \infty$ to $- \infty$ $(C_2)$. The real time 
formalism is achieved by doubling every degree of freedom, 
$\vec{\phi} \to (\vec{\phi}_1, \vec{\phi}_2)$ and $\vec{\varphi} \to 
(\vec{\varphi}_1, \vec{\varphi}_2)$, where $\phi_1 (x_0, {\bf x}) = 
\phi (x_0, {\bf x})$ with $x_0 \in C_1$ and $\phi_2 (x_0, {\bf x}) = 
\phi (x_0, {\bf x})$ with $x_0 \in C_2$, etc. A classical contour 
action is written in the form 
\begin{eqnarray} 
&& \int_C d x_0 \int d {\bf x} \, {\cal L} (\vec{\phi} (x); 
\vec{\varphi} (x)) = \int_{- \infty}^{+ \infty} d x_0 \int 
d {\bf x} \, \hat{\cal L} (x) \, , \nonumber \\ 
&& \mbox{\hspace*{5ex}} \hat{\cal L} \equiv {\cal L} (\vec{\phi}_1; 
\vec{\varphi}_1) - {\cal L} (\vec{\phi}_2; \vec{\varphi}_2) \, . 
\label{hat-L} 
\end{eqnarray} 
$\hat{\cal L}$ here is sometimes called a hat-Lagrangian.  

Taking $\hat{\cal L}_0$, which corresponds to ${\cal L}_0$ in 
Eq.~(\ref{2.8}), for a free hat-Lagrangian, we construct from first 
principles a perturbative framework. Throughout this paper, we do 
not deal with initial correlations (see, e.g., \cite{chou}). 
Following standard procedure, the four kind of propagators emerges: 
\begin{eqnarray} 
& \Delta_{1 1}^{\alpha \beta} (x, y) = - i \mbox{Tr} \left[ T 
\left( \phi_1^\alpha (x) \phi_1^\beta (y) \right) \, \rho 
\right] \, , \;\;\;\;\;\; & \Delta_{2 2}^{\alpha \beta} (x, y) = - 
i \mbox{Tr} \left[ \overline{T} \left( \phi_2^\alpha (x) 
\phi_2^\beta (y) \right) \, \rho \right] \, , \nonumber \\ 
& \Delta_{1 2}^{\alpha \beta} (x, y) = - i \mbox{Tr} \left[ 
\phi_2^\beta (y) \phi_1^\alpha (x) \, \rho \right] \, , 
\;\;\;\;\;\; & \Delta_{2 1}^{\alpha \beta} (x, y) = - i \mbox{Tr} 
\left[ \phi_2^\alpha (x) \phi_1^\beta (y) \, \rho \right] \, , 
\label{den} 
\end{eqnarray} 
where $\rho$ is the density matrix, $T$ is a time-ordering symbol, 
and $\overline{T}$ is an anti-time-ordering symbol. At the end of 
calculation we set $\vec{\phi}_1 = \vec{\phi}_2$ and 
$\vec{\varphi}_1 = \vec{\varphi}_2$ \cite{chou}. Let us introduce a 
matrix propagator $\hat{\bf \Delta} (x, y)$, where the bold-face 
denotes, as above, the $N \times N$ matrix and the \lq caret' 
denotes the $2 \times 2$ matrix: $\Delta_{i j}^{\alpha \beta}$ is 
the $(\alpha, \beta)$-component of the $N \times N$ matrix ${\bf 
\Delta}_{i j}$ and, at the same time, $(i, j)$-component of the $2 
\times 2$ matrix $\hat{\Delta}^{\alpha \beta}$. The matrix 
self-energy part $\hat{\bf \Sigma} (x, y)$ is defined similarly. 

We stress again that the argument \lq $x$' of ${\bf P}_\xi (x)$ 
$(\xi = \pi, \sigma)$ in Eq.~(\ref{inv}) is macroscopic spacetime 
coordinates. Let $\hat{\bf A}$ be a propagator or a self-energy 
part. Due to $O (N - 1)$ symmetry of ${\cal L}$, Eq.~(\ref{owari}), 
we may write $\hat{\bf A}$ as 
\begin{equation} 
\hat{\bf A} (x, y) \simeq {\bf P}_\xi (x) \hat{A}^{(\xi)} (x, y) 
{\bf P}_\xi (y) + {\bf T}_{\xi \underline{\xi}}^\mu (x) 
\hat{A}^{(\xi)}_\mu (x, y) \, , 
\label{kihon} 
\end{equation} 
where, for $\xi = \pi$ $(\sigma)$, $\underline{\xi} = \sigma$ 
$(\pi)$ and 
\begin{equation} 
{\bf T}_{\sigma \pi}^\mu (x) \equiv | \hat{\varphi} (x) \rangle 
\partial_{x_\mu} \langle \hat{\varphi} (x) | \, , \;\;\;\;\;\; 
{\bf T}_{\pi \sigma}^\mu (x) \equiv  \hat{\varphi} (x) \rangle 
\stackrel{\leftarrow}{\partial}_{x_\mu} \langle \hat{\varphi} (x) | 
\label{t} 
\end{equation} 
(cf. Sec.~III). One can replace ${\bf T}^\mu_{\xi \underline{\xi}} 
(x)$ in Eq.~(\ref{kihon}) with ${\bf T}^\mu_{\xi \underline{\xi}} 
(y)$, since the arising difference is of higher order (cf. 
Eq.~(\ref{gra})). Fourier transforming $\hat{A}^{(\xi)}$ and 
$\hat{A}^{(\xi)}_\mu$ with respect to $x - y$, we have 
\begin{equation} 
\hat{\bf A} (x, y) \simeq {\bf P}_\xi (x) \int \frac{d^{\, 4} 
P}{(2 \pi)^4} e^{- i P \cdot (x - y)} \, \hat{A}^{(\xi)} (X; P) {\bf 
P}_\xi (y) + {\bf T}_{\xi \underline{\xi}}^\mu (x) \int \frac{d^{\, 
4} P}{(2 \pi)^4} e^{- i P \cdot (x - y)} \, \hat{A}^{(\xi)}_\mu (X; 
P) \, , 
\label{kihon1} 
\end{equation} 
where $X \equiv (x + y) / 2$. In general, $\hat{A}^{(\xi)}$ in 
Eq.~(\ref{kihon1}) consists of two pieces, $\hat{A}^{(\xi)} = 
\hat{A}^{(\xi)}_0 + \hat{A}^{(\xi)}_1$, where $\hat{A}^{(\xi)}_0$ is 
free from $X$-derivative and $\hat{A}^{(\xi)}_1$ $(= 
\hat{A}^{(\xi)}_1 (X; P))$ contains explicit (first order) 
$X_\mu$-derivative: 
\begin{eqnarray} 
\hat{\bf A} (x, y) & \simeq & {\bf P}_\xi (x) \int \frac{d^{\, 4} 
P}{(2 \pi)^4} e^{- i P \cdot (x - y)} \, \hat{A}^{(\xi)}_0 (X; P) 
{\bf P}_\xi (y) + {\bf P}_\xi (x) \int \frac{d^{\, 4} 
P}{(2 \pi)^4} e^{- i P \cdot (x - y)} \, \hat{A}^{(\xi)}_1 (X; P) 
{\bf P}_\xi (y) \nonumber \\ 
&& + {\bf T}_{\xi \underline{\xi}}^\mu (x) \int \frac{d^{\, 4} P}{(2 
\pi)^4} e^{- i P \cdot (x - y)} \, \hat{A}^{(\xi)}_\mu (X; P) \, . 
\label{kihon2} 
\end{eqnarray} 
The first term on the R.H.S. is the leading part of $\hat{\bf A}$ 
while the second and third terms are the gradient parts (cf. above 
after Eq.~(\ref{gra})). 
\subsection{Propagator} 
From the definition of $\Delta$'s, Eq.~(\ref{den}), with 
$\vec{\bf \phi}_1 = \vec{\bf \phi}_2$ (cf. above after 
Eq.~(\ref{den})), we see that 
\begin{equation} 
\sum_{i, \, j = 1}^2 (-)^{i + j} {\bf \Delta}_{i j} 
\, \rule[-4mm]{.14mm}{9.5mm} \raisebox{-3.85mm}{\scriptsize{$\; 
\vec{\phi}_1 = \vec{\phi}_2$}} = 0 
\label{wa} 
\end{equation} 
holds. Then, out of four ${\bf \Delta}_{i j}$ $(i, j = 1, 2)$, three 
are independent, for which we choose \cite{chou} 
\[ 
{\bf \Delta}_R = {\bf \Delta}_{1 1} - {\bf \Delta}_{1 2} \, , 
\;\;\;\;\; {\bf \Delta}_A = {\bf \Delta}_{1 1} - {\bf \Delta}_{2 1} 
\, , \;\;\;\;\; {\bf \Delta}_c = {\bf \Delta}_{1 2} + {\bf 
\Delta}_{2 1} \, . 
\] 
Setting $\vec{\phi}_1 = \vec{\phi}_2$ $(\equiv \vec{\phi})$, we have 
\begin{eqnarray} 
i \Delta_R^{\alpha \beta} (x, y) & = & \theta (x_0 - y_0) \mbox{Tr} 
\left\{ \left[ \phi^\alpha (x), \, \phi^\beta (y) \right] \, \rho 
\right\} = \theta (x_0 - y_0) \left[ \phi^\alpha (x), \, \phi^\beta 
(y) \right] \, , \nonumber \\ 
i \Delta_A^{\alpha \beta} (x, y) & = & - \theta (y_0 - x_0) 
\mbox{Tr} \left\{ \left[ \phi^\alpha (x), \, \phi^\beta (y) \right] 
\, \rho \right\} = - \theta (y_0 - x_0) \left[ \phi^\alpha (x), \, 
\phi^\beta (y) \right] \, , \nonumber \\ 
i \Delta_c^{\alpha \beta} (x, y) & = & - i \mbox{Tr} \left[ \left( 
\phi^\alpha (x) \, \phi^\beta (y) + \phi^\beta (y) \, \phi^\alpha 
(x) \right) \, \rho \right] \, . \nonumber \\ 
\label{doku1} 
\end{eqnarray} 
Thus, ${\bf \Delta}_R$ and  ${\bf \Delta}_A$ are the retarded- and 
advanced-propagators, respectively, which have already been 
constructed in Sec.~III. ${\bf \Delta}_c$ is the correlation 
function. Expressing $\hat{\bf \Delta}$ in terms of them, we have 
\begin{equation} 
\hat{\bf \Delta} = \frac{1}{2} \left( 
\begin{array} {cc} 
{\bf \Delta}_R + {\bf \Delta}_A & \;\; - {\bf \Delta}_R + 
{\bf \Delta}_A \\ 
{\bf \Delta}_R - {\bf \Delta}_A & \;\; - {\bf \Delta}_R - 
{\bf \Delta}_A 
\end{array} 
\right) 
+ \frac{1}{2} {\bf \Delta}_c \hat{A}_+ \, , 
\label{prop} 
\end{equation} 
where 
\begin{equation} 
\hat{A}_\pm = \left( 
\begin{array} {cc} 
1 & \; \pm 1 \\ 
\pm 1 & \; 1 
\end{array} 
\right) \, . 
\label{mat} 
\end{equation} 

From the definition of ${\bf \Delta}_c$ and Eq.~(\ref{den}) with 
$\vec{\bf \phi}_1 = \vec{\bf \phi}_2$, it follows that 
\begin{equation} 
\left( i {\bf \Delta}_c (x, y) \right)^* =  i {\bf \Delta}_c (x, 
y) \, , \;\;\;\;\;\; \Delta_c^{\alpha \beta} (x, y)  =  
\Delta_c^{\beta \alpha} (y, x) \, . 
\label{seisitu} 
\end{equation} 
To leading order (of derivative expansion), it is known that ${\bf 
\Delta}_c$ takes the form (cf., e.g., \cite{nie}) 
\[ 
{\bf \Delta}_c (x, y) \sim {\bf P}_\xi (x) \int \frac{d^{\, 4} 
P}{(2 \pi)^4} \, e^{- i P \cdot (x - y)} \left( 1 + 2 f_\xi (X; P) 
\right) \left( \Delta_R^{(\xi)} (X; P) - 
\Delta_A^{(\xi)} (X; P) \right) {\bf P}_\xi (y) \, , 
\] 
where $X \equiv (x + y) / 2$ and, as will be seen below, $f_\xi$ is 
the real function that is related to the particle-number density. 
Then, to the gradient approximation, we may write 
\begin{equation} 
{\bf \Delta}_c (x, y) = {\bf P}_\xi (x) \Delta_c^{(\xi)} (x, y) 
{\bf P}_\xi (y) + 
{\bf T}_{\xi \underline{\xi}}^\mu (x) 
\Delta_\mu^{(\xi)} (x, y) \, , 
\label{paramet} 
\end{equation} 
where, ${\bf T}_{\xi \underline{\xi}}^\mu (x)$ is as in 
Eq.~(\ref{t}) and 
\begin{equation} 
\Delta_c^{(\xi)} = \Delta_R^{(\xi)} \cdot (1 + 2 f_\xi) - (1 + 
2 f_\xi) \cdot \Delta_A^{(\xi)} + \Delta_{c 1}^{(\xi)} \, , 
\label{1st} 
\end{equation} 
where $\xi$ stands for $\pi$ or $\sigma$, and $\Delta_{c 1}^{(\xi)}$ 
is a gradient part. It is clear that, in Eq.~(\ref{1st}), although 
$\xi$ is a \lq repeated index' on the R.H.S., summation should not 
be taken over $\xi$. This type of equations appears frequently in 
the sequel. Here we have used the short-hand notation $F \cdot 
G$, which is a function whose \lq\lq $(x, y)$-component'' is 
\begin{equation} 
[F \cdot G] (x, y) = \int d^{\, 4} z \, F (x, z) G (z, y) \, , 
\label{ab} 
\end{equation} 
with \lq $1$' in Eq.~(\ref{1st}) the function whose $(x, 
y)$-com\-po\-nent is $\delta^4 (x - y)$. For a given $\rho$, ${\bf 
\Delta}_c$ is computed through Eq.~(\ref{doku1}). 
Eq.~(\ref{paramet}) with Eq.~(\ref{1st}) is understood to be the 
defining equation of $f_\xi$ and $\Delta_{c 1}^{(\xi)}$. Physical 
meaning of $f_\xi$ is clarified later (see, Sec.~VIC). Using 
Eq.~(\ref{seisitu}) in Eq.~(\ref{paramet}) with Eq.~(\ref{1st}), we 
obtain the relations: 
\begin{eqnarray*} 
& \left( \Delta_c^{(\xi)} (x, y) \right)^* = - \Delta_c^{(\xi)} 
(x, y) \, , \;\;\;\;\;\; & \left( \Delta_\mu^{(\xi)} (x, y) \right)^* 
= - \Delta_\mu^{(\xi)} (x, y) \, , \\ 
& \Delta_c^{(\xi)} (x, y) = \Delta_c^{(\xi)} (y, x) \, , 
\;\;\;\;\;\; & \Delta_\mu^{(\xi)} (x, y) = 
\Delta_\mu^{(\underline{\xi})} (y, x) \, . 
\end{eqnarray*} 
Note that ${\bf T}_{\xi \underline{\xi}}^\mu \Delta_\mu^{(\xi)}$ in 
Eq.~(\ref{paramet}) and $\Delta_{c 1}^{(\xi)}$ in Eq.~(\ref{1st}) 
are the gradient parts (cf. above after Eq.~(\ref{kihon2})). 

Applying $\hat{\tau}_3 (\partial^2 + {\cal M}_\xi^2) {\bf P}_\xi$ 
($= \hat{\tau}_3 {\bf P}_\xi (\partial^2 + {\cal M}_\xi^2)$) to 
${\bf \Delta}_c \hat{A}_+$, we obtain 
\begin{eqnarray} 
&& \hat{\tau}_3 \left( \partial^2_x + {\cal M}_\xi^2 (x) \right) 
{\bf P}_\xi (x) {\bf \Delta}_c (x, y) \hat{A}_+ \nonumber \\ 
&& \mbox{\hspace*{3ex}} \simeq {\bf P}_\xi (X) \int \frac{d^{\, 4} 
P}{(2 \pi)^4} e^{- i P \cdot (x - y)} \left[ 2 i \left\{ f_\xi, \, 
P^2 - {\cal M}_\xi^2 \right\} \Delta_A^{(\xi)} (X; P) \right. 
\nonumber \\ 
& & \mbox{\hspace*{6ex}} + 2 i \left( P \cdot \partial {\bf P}_\xi 
(X) \right) \left( 1 + 2 f_{\underline{\xi}} (X; P) \right) \left( 
\Delta_R^{(\underline{\xi})} (X; P) - \Delta_A^{(\underline{\xi})} 
(X: P) \right) \nonumber \\ 
&& \mbox{\hspace*{6ex}} \left. - \left( P^2 - {\cal M}_\xi^2 (X) 
\right) \left( \Delta_{c 1}^{(\xi)} (X; P) + {\bf T}_{\xi 
{\underline{\xi}}}^\mu (X) \Delta_\mu^{(\xi)} (X; P) \right) \right] 
\hat{\tau}_3 \hat{A}_+ \, , 
\label{4.14} \\ 
&& \hat{A}_+ {\bf \Delta}_c (x, y) 
(\stackrel{\leftarrow}{\partial}^2_y + {\cal M}_\xi^2 (y)) {\bf 
P}_\xi (y) \hat{\tau}_3 \nonumber \\ 
& & \mbox{\hspace*{3ex}} \simeq {\bf P}_\xi (X) \int \frac{d^{\, 4} 
P}{(2 \pi)^4} e^{- i P \cdot (x - y)} \left[ 2 i \left\{ f_\xi (X; 
P) , \, P^2 - {\cal M}_\xi^2 (X) \right\} \Delta_R^{(\xi)} (X; P) 
\right. \nonumber \\ 
& & \mbox{\hspace*{6ex}} + 2 i \left( P \cdot \partial {\bf P}_\xi 
(X) \right) \left( 1 + 2 f_\xi (X; P) \right) \left( 
\Delta_R^{(\xi)} (X; P) - \Delta_A^{(\xi)} (X; P) \right) \nonumber 
\\ 
& & \mbox{\hspace*{6ex}} \left. - \left( P^2 - {\cal M}_\xi^2 (X) 
\right) \Delta_{c 1}^{(\xi)} (X; P) - \left( P^2 - 
{\cal M}_{\underline{\xi}}^2 (X) \right) 
{\bf T}_{\xi {\underline{\xi}}}^\mu (X) \Delta_\mu^{(\xi)} (X; P) 
\right] \hat{A}_+ 
\hat{\tau}_3 \, , 
\label{enzan} 
\end{eqnarray} 
where 
\begin{eqnarray*} 
\left\{ f_\xi (X; P) , \, P^2 - {\cal M}_\xi^2 (X) \right\} & \equiv 
& \frac{\partial f_\xi (X; P)}{\partial X_\mu} 
\frac{\partial (P^2 - {\cal M}_\xi^2 (X))}{\partial P^\mu} 
- \frac{\partial f_\xi (X; P)}{\partial P^\mu} 
\frac{\partial (P^2 - {\cal M}_\xi^2 (X))}{\partial X_\mu} \\ 
&& = 2 P \cdot \partial_X f_\xi (X; P) + 
\frac{\partial f_\xi (X; P)}{\partial P^\mu} 
\frac{\partial {\cal M}_\xi^2 (X)}{\partial X_\mu} \, . 
\end{eqnarray*} 
\subsubsection*{Bare-$N$ scheme} 
The propagator matrix $\hat{\bf \Delta}$ is an inverse of $- 
\hat{\tau}_3 (\partial^2 + {\cal M}_\xi^2) {\bf P}_\xi$ (cf. 
Eq.~(\ref{hat-L}) and Eq.~(\ref{2.8}) with Eq.~(\ref{inv})). Then, 
Eqs.~(\ref{4.14}) and (\ref{enzan}) should vanish. From this 
condition, we obtain the following relations: 
\begin{eqnarray} 
&& \left\{ f_\xi, \, P^2 - {\cal M}_\xi^2 \right\} = 0 \, , 
\label{free-B} 
\\ 
&& (P^2 - {\cal M}_\xi^2 (X)) \Delta_{c 1}^{(\xi)} (X; P) = 0  \, , 
\label{cond1} 
\\ 
&& \Delta_\mu^{(\sigma)} (X; P) = - \Delta_\mu^{(\pi)} (X; P) 
\nonumber \\ 
&& \mbox{\hspace*{11ex}} = \frac{2 i P_\mu}{{\cal M}_\sigma^2 (X) 
- {\cal M}_\pi^2 (X)} \left[ \left( 1 + 2 f_\sigma (X; P) \right) 
\left( \Delta_R^{(\sigma)} (X; P) - \Delta_A^{(\sigma)} (X; P) 
\right) \right. \nonumber \\ 
&& \mbox{\hspace*{13.5ex}} \left. - \left( 1 + 2 f_\pi (X; P) 
\right) \left( \Delta_R^{(\pi)} (X; P) - \Delta_A^{(\pi)} (X; P) 
\right) \right] \, . 
\label{cond2} 
\end{eqnarray} 

\noindent 
As will be shown later (cf. Eq.~(\ref{kank})), $f_\xi$ $(\xi = \pi, 
\sigma)$ is related to the number density $N_\xi$ of $\xi$: $f_\xi 
(X; \tau E_p^{(\xi)}, \hat{\bf p}) = - \theta (- \tau) + \epsilon 
(\tau) N_\xi (X; E_p^{(\xi)}, \epsilon (\tau) \hat{\bf p})$ $(\tau = 
\pm)$. Then, Eq. (\ref{free-B}) is a \lq\lq free Boltzmann 
equation.'' One can construct a perturbation theory in a similar 
manner as in \cite{nie}, where a complex-scalar field system with 
symmetric phase is treated. We call the perturbation theory thus 
constructed the bare-$N$ scheme, since $N_\xi$ obeys the \lq\lq free 
Boltzmann equation.'' This theory is equivalent \cite{nie} to the 
one obtained in the physical-$N$ scheme, to which we now turn. 
\subsubsection*{Physical-$N$ scheme} 
We abandon Eq.~(\ref{free-B}), while we keep Eqs.~(\ref{cond1}) and 
(\ref{cond2}). This means that $f_\xi$ in the present (physical-$N$) 
scheme differs from $f_\xi$ in the bare-$N$ scheme. Specification of 
$f_\xi$ is postponed until Sec.~VIC, where we require the number 
density $N_\xi$ to be as close as possible to the physical number 
density. Now, $\hat{\bf \Delta}$ is not an inverse of $- 
\hat{\tau}_3 (\partial^2 + {\cal M}_\xi^2) {\bf P}_\xi$. It is 
straightforward to show in the gradient approximation that $\hat{\bf 
\Delta}$ is an inverse of $- \hat{\tau}_3 (\partial^2 + {\cal 
M}_\xi^2) {\bf P}_\xi + {\bf L}' (x, y) \hat{A}_-$, where 
$\hat{A}_-$ is as in Eq.~(\ref{mat}) and 
\begin{equation} 
{\bf L}' (x, y) = i {\bf P}_\xi (X) \int \frac{d^{\, 4} P}{(2 
\pi)^4} e^{- i P \cdot (x - y)} \left\{ f_\xi (X; P) , \; P^2 - 
{\cal M}_\xi^2 (X) \right\} \, . 
\label{L-p} 
\end{equation} 
Then the free action is 
\begin{equation} 
{\cal A}_0 = - \frac{1}{2} \int d^{\, 4} x \, \hat{\phi} (x) 
\hat{\tau}_3 \left( \partial^2 + {\cal M}_\xi^2 (x) {\bf P}_\xi (x) 
\right) \hat{\phi} (x) + \frac{1}{2} \int d^{\, 4} x \, d^{\, 4} y 
\, \hat{\phi} (x) {\bf L}' (x, y) \hat{A}_- \hat{\phi} (y) \, . 
\label{act1} 
\end{equation} 
Note that the Lagrangian density corresponding to the last term of 
Eq.~(\ref{act1}) is nonlocal not only in \lq space' but also in \lq 
time.' Here it is worth mentioning the so-called 
$|p_0|$-prescription. With this prescription, at an intermediate 
stage, we have ${\bf L}' (x, y)$, which is local in time. (For 
completeness, we briefly discuss the $|p_0|$-prescription in 
Appendix A.) 

Since the last term of Eq.~(\ref{act1}) is absent in the original 
action, we should introduce the counter action to compensate it, 
\begin{equation} 
{\cal A}_c = - \frac{1}{2} \int d^{\, 4} x \, d^{\, 4} y \hat{\phi} 
(x) {\bf L}' (x, y) \hat{A}_- \hat{\phi} (y) \, , 
\label{act} 
\end{equation} 
which yields a vertex $- i {\bf L}' \hat{A}_-$ 
$(\equiv i \hat{\bf V}_c)$. From Eqs.~(\ref{prop}) and (\ref{act}), 
$\hat{\bf V}_c \cdot \hat{\bf \Delta} \cdot \hat{\bf V}_c = 0$ 
follows, and the $\hat{\bf V}_c$-resummed propagator becomes 
\[
\hat{\bf \Delta}_{c-resum} = \hat{\bf \Delta} \cdot \left[ 
1 + \sum_{n = 1}^\infty \left( - \hat{\bf V}_c \cdot \hat{\bf 
\Delta} \right)^n \right] =  \hat{\bf \Delta} - \hat{\bf \Delta} 
\cdot \hat{\bf V}_c \cdot \hat{\bf \Delta} \; \left( \equiv 
\hat{\bf \Delta} + \delta \hat{\bf \Delta} \right) \, . 
\] 
Since $\hat{\bf V}_c$ is a gradient part, we obtain, to the 
gradient approximation, 
\begin{equation} 
\delta \hat{\bf \Delta} \simeq i {\bf P}_\xi (X) \int 
\frac{d^{\, 4} P}{(2 \pi)^4} e^{- i P \cdot (x - y)} \left\{ f_\xi, 
\; P^2 - {\cal M}_\xi^2 \right\} \Delta_R^{(\xi)} \Delta_A^{(\xi)} 
\hat{A}_+ \, . 
\label{16} 
\end{equation} 
Note that Eq.~(\ref{16}) possesses pinch singularities in a 
$p_0$-plane, due to $\Delta_R^{(\xi)} (X; P) \Delta_A^{(\xi)} (X; 
P)$. Since $\delta \hat{\bf \Delta}$ is proportional to $\hat{A}_+$, 
it contributes to ${\bf \Delta}_c$ (cf. Eq.~(\ref{prop})). Then, 
including $\delta \hat{\bf \Delta}$ to ${\bf \Delta}_c \hat{A}_+ / 
2$, we obtain for ${\bf \Delta}_c$ (within the gradient 
approximation), 
\begin{eqnarray} 
{\bf \Delta}_c (x, y) & = & - i {\bf P}_\xi (x) \int \frac{d^{\, 4} 
P}{(2 \pi)^3} e^{- i P \cdot (x - y)} \epsilon (p_0) \left( 1 + 2 
f_\xi (X; P) \right) \delta (P^2 - {\cal M}_\xi^2 (X)) {\bf P}_\xi 
(y) \nonumber \\ 
&& + \int \frac{d^{\, 4} P}{(2 \pi)^4} e^{- i P \cdot (x - y)} 
\left[ - i {\bf P}_\xi (X) \left\{ f_\xi, \; P^2 - {\cal M}_\xi^2 
\right\} \left( \Delta_R^{(\xi)} (X; P) - \Delta_A^{(\xi)} (X; P) 
\right)^2 
\right. \nonumber \\ 
&& \mbox{\hspace*{5ex}} \left. + {\bf P}_\xi (X) \Delta_{c 
1}^{(\xi)} (X; P) + \left( | \hat{\varphi} (X) 
\rangle \stackrel{\leftrightarrow}{\partial}_{X_\mu} \langle 
\hat{\varphi} 
(X) | \right) \Delta_\mu^{(\sigma)} (X; P) \right] \, , 
\label{cc} 
\end{eqnarray} 
where $\Delta_\mu^{(\sigma)}$ is as in Eq.~(\ref{cond2}). From 
Eq.~(\ref{cond1}), we see that $\Delta_{c 1}^{(\xi)} (X; P) \propto 
\delta (P^2 - {\cal M}_\xi^2 (X))$, so that the term with $\Delta_{c 
1}^{(\xi)}$ in Eq.~(\ref{cc}) may be absorbed into the first term of 
the R.H.S. by modifying the definition of $f_\xi (X; P)$. Thus we 
shall drop the term with $\Delta_{c 1}^{(\xi)}$ hereafter. 

Substituting Eqs.~(\ref{RA-de}), (\ref{ana}), and (\ref{cc}) into 
Eq.~(\ref{prop}), we obtain 
\begin{eqnarray} 
\hat{\bf \Delta} (x, y) & = & \hat{\bf \Delta}^{(0)} (x, y) + 
\int \frac{d^{\, 4} P}{(2 \pi)^4} e^{ - i P \cdot (x - y)} \left[ 
\hat{\bf \Delta}^{(p)} (X; P) + \hat{\bf \Delta}^{(t)} (X; P) 
\right] \, , 
\label{ato} \\ 
\hat{\bf \Delta}^{(0)} (x, y) & = & {\bf P}_\xi (x) \int \frac{d^{\, 
4} P}{(2 \pi)^4} \, e^{- i P \cdot (x - y)} 
\hat{\Delta}^{(\xi)} (X; P) {\bf P}_\xi (y) \, , \nonumber \\ 
\hat{\bf \Delta}^{(p)} (X; P) & = & - \frac{i}{2} {\bf P}_\xi (X) 
\left\{ f_\xi, \, P^2 - {\cal M}_\xi^2 \right\} \left( 
\Delta_R^{(\xi)} - \Delta_A^{(\xi)} \right)^2 \hat{A}_+ \, , 
\nonumber \\ 
\hat{\bf \Delta}^{(t)} (X; P) & = & 2 i | \hat{\varphi} \rangle P 
\cdot \stackrel{\leftrightarrow}{\partial} \langle \hat{\varphi}| 
\hat{\Omega} \, , 
\label{prop-fin} 
\end{eqnarray} 
where $\Delta_R^{(\xi)} = \Delta_R^{(\xi)} (X; P)$, $f_\xi = f_\xi 
(X; P)$, etc., and 
\begin{eqnarray} 
\hat{\Delta}^{(\xi)} (X; P) & = & \left( 
\begin{array}{cc} 
\Delta_R^{(\xi)} + f_\xi \left( \Delta_R^{(\xi)} - \Delta_A^{(\xi)} 
\right) & \;\;\; f_\xi \left( \Delta_R^{(\xi)} - \Delta_A^{(\xi)} 
\right) \\ 
\left( 1 + f_\xi \right) \left( \Delta_R^{(\xi)} - \Delta_A^{(\xi)} 
\right) & \;\;\; - \Delta_A^{(\xi)} + f_\xi \left( \Delta_R^{(\xi)} 
- \Delta_A^{(\xi)} \right) 
\end{array} 
\right) \, , 
\label{lead} \\ 
\hat{\Omega} & = & \left( 
\begin{array}{cc} 
\Delta_R^{(\pi)} \Delta_R^{(\sigma)} \, , & \;\;\; 0 
\\ 
\Delta_R^{(\pi)} \Delta_R^{(\sigma)} - \Delta_A^{(\pi)} 
\Delta_A^{(\sigma)} \, , & \;\;\; - \Delta_A^{(\pi)} 
\Delta_A^{(\sigma)} 
\end{array} 
\right) + \omega \hat{A}_+ \, , \nonumber \\ 
\omega & \equiv & \frac{1}{{\cal M}_\sigma^2 - {\cal M}_\pi^2} 
\left[ f_\sigma \left( \Delta_R^{(\sigma)} - \Delta_A^{(\sigma)} 
\right) - f_\pi \left( \Delta_R^{(\pi)} - \Delta_A^{(\pi)} \right) 
\right] \, . \nonumber 
\end{eqnarray} 
As has been observed in Sec.~III, $|\hat{\varphi} \rangle P \cdot 
\partial \langle \hat{\varphi}|$ and $| \hat{\varphi} \rangle P 
\cdot \stackrel{\leftarrow}{\partial} \langle \hat{\varphi}|$ in the 
gradient part $\hat{\bf \Delta}^{(t)}$ \lq\lq induce'' transition 
between $\pi$ mode and $\sigma$ mode. Thus, $\hat{\bf \Delta}^{(t)}$ 
contributes, e.g., to the one-loop two-point function with one-$\pi$ 
and one-$\sigma$ legs.  
\subsection{Feynman rules} 
Two fundamental elements of Feynman rules are the propagator and the 
vertices, which take the $(2 \times 2)$ matrix form. The 
propagator-matrix is given by Eq.~(\ref{ato}). The vertex factors 
may be reads off from Eq.~(\ref{hat-L}) with Eqs.~(\ref{owari}), 
(\ref{L-q}), (\ref{2.6d}) and (\ref{teigi-yo}). As a matter of fact, 
from Eq.~(\ref{hat-L}), it is obvious that there is no vertex that 
mixes $\phi_1$'s with $\phi_2$'s. The vertex factors for the fields 
$\phi_1$'s are the same as in vacuum theory. Each vertex factor for 
$\phi_2$'s is of opposite sign to the corresponding vertex factor 
for $\phi_1$'s. Thus, in matrix notation as for the propagator, 
every vertex-matrix $\hat{\bf V}$ is diagonal, $\hat{\bf V} = 
\mbox{diag.} ({\bf v}, - {\bf v})$, with ${\bf v}$ the same as in 
vacuum theory. All other elements of Feynman rules, e.g., 
integration over every loop momentum, are the same as in vacuum 
theory. 

Having thus constructed Feynman rules, one can compute reaction 
rates for processes taking place in the system, by using the 
reaction-rate formula \cite{oka}. 
\setcounter{equation}{0}
\setcounter{section}{4} 
\def\theequation{\mbox{\arabic{section}.\arabic{equation}}} 
\section{Quasiparticle representation of the propagator} 
Here we obtain a quasiparticle representation \cite{ume} of 
$\hat{\bf \Delta}$, which tremendously simplifies the practical 
computation.  Straightforward but lengthy calculation shows that 
$\hat{\bf \Delta}$ may be written in the form (cf. the definition 
(\ref{ab})): 
\begin{equation} 
\hat{\bf \Delta} \simeq \hat{\bf B}_L \cdot \left( 
\begin{array}{cc} 
{\bf \Delta}_R & \;\; 0 \\ 
0 & \;\; - {\bf \Delta}_A 
\end{array} 
\right) \cdot \hat{\bf B}_R \, , 
\label{BB} 
\end{equation} 
where 
\begin{eqnarray*} 
\hat{\bf B}_L & \simeq & \hat{\bf B}_L^{(0)} + | \hat{\varphi} (X) 
\rangle \stackrel{\leftrightarrow}{\partial}_{X^\mu} \langle 
\hat{\varphi} (X)| \alpha^\mu (X) \left( 
\begin{array}{cc} 
0 & \;\; 1 \\ 
0 & \;\; 1 
\end{array} \right) \, , \\ 
\hat{\bf B}_R & \simeq & \hat{\bf B}_R^{(0)} + | \hat{\varphi} (X) 
\rangle \stackrel{\leftrightarrow}{\partial}_{X^\mu} \langle 
\hat{\varphi} (X)| \alpha^\mu (X) \left( 
\begin{array}{cc} 
1 & \;\; 1 \\ 
0 & \;\; 0 
\end{array} \right) \, , 
\end{eqnarray*} 
with 
\begin{eqnarray} 
\hat{\bf B}_L^{(0)} & = & {\bf P}_\xi \cdot \left( 
\begin{array}{cc} 
1 & \;\; f_\xi \\ 
1 & \;\; 1 + f_\xi 
\end{array} 
\right) \cdot {\bf P}_\xi \, , \;\;\;\;\;\;\;\;\;\;\; 
\hat{\bf B}_R^{(0)} = {\bf P}_\xi \cdot \left( 
\begin{array}{cc} 
1 + f_\xi & \;\; f_\xi \\ 
1 & \;\; 1 
\end{array} 
\right) \cdot {\bf P}_\xi \, , 
\label{B00} \\ 
\alpha^\mu (X) & = & \frac{2 i}{{\cal M}_\sigma^2 (X) - {\cal 
M}_\pi^2 (X)} \int \frac{d^{\, 4} P}{(2 \pi)^4} \, e^{- i P \cdot 
(x - y)} P^\mu \left( f_\sigma (X; P) - f_\pi (X; 
P) \right) \, . \nonumber 
\end{eqnarray} 
It can readily be shown that 
\begin{equation} 
\hat{\bf B}_L \cdot \hat{\tau}_3 \hat{\bf B}_R \simeq 
\hat{\bf B}_R \cdot \hat{\tau}_3 \hat{\bf B}_L \simeq \hat{\tau}_3 
\, . 
\label{6.5} 
\end{equation} 
It is obvious from the argument in Appendix A that, at an 
intermediate stage before taking the $|p_0|$-prescription, $\hat{\bf 
B}_L$ and $\hat{\bf B}_R$ are local in time and satisfy the relation 
(\ref{6.5}). Then, $\hat{\bf B}_L$ and $\hat{\bf B}_R$ at that stage 
are the generalized Bogoliubov-matrices \cite{ume}. Eq.~(\ref{BB}) 
tells us that, through Bogoliubov-transforming $\vec{\phi}_i$ $(i = 
1, 2)$, one can introduce the stable quasiparticle modes, 
$\vec{\phi}_i'$ $(i = 1, 2)$, whose matrix-propagator is 
$\mbox{diag}. ({\bf \Delta}_R, - {\bf \Delta}_A)$ in Eq.~(\ref{BB}) 
(cf. \cite{nie} for more details). 

It is worth mentioning that $\hat{\bf B}_L$ and $\hat{\bf B}_R$ may 
be written as 
\begin{eqnarray*} 
&& \hat{\bf B}_L (x, y) \simeq {\bf P}_\xi (x) \left( 
\begin{array}{cc} 
\delta^{\, 4} (x - y) & \;\; F_\xi (x, y) \\ 
\delta^{\, 4} (x - y) & \;\; \delta^{\, 4} (x - y) + F_\xi (x, y) 
\end{array} 
\right) {\bf P}_\xi (y) \, , 
\\ 
&& \hat{\bf B}_R (x, y) \simeq {\bf P}_\xi (x) \left( 
\begin{array}{cc} 
\delta^{\, 4} (x - y) + F_\xi (x, y) & \;\; F_\xi (x, y) \\ 
\delta^{\, 4} (x - y) & \;\; \delta^{\, 4} (x - y) 
\end{array} 
\right) {\bf P}_\xi (y) \, , 
\end{eqnarray*} 
where 
\[ 
F_\xi (x, y) = f_\xi (x, y) + 2 
\frac{\stackrel{\leftarrow}{\partial}_x  
\stackrel{\rightarrow}{\partial}_x f_\xi (x, y) 
+ f_\xi (x, y) \stackrel{\leftarrow}{\partial}_y 
\stackrel{\rightarrow}{\partial_y}}{{\cal M}_\xi^2 (X) - {\cal 
M}_{\underline{\xi}}^2 (X)} 
\] 
with $\underline{\pi} = \sigma$ and $\underline{\sigma} = \pi$. 
\setcounter{equation}{0}
\setcounter{section}{5} 
\def\theequation{\mbox{\arabic{section}.\arabic{equation}}} 
\section{The gap equation and the generalized Boltzmann equation} 
\subsection{Self-energy-part resummed propagator} 
We write the bare propagator $\hat{\bf \Delta}$ in Eq.~(\ref{ato}) 
as $\hat{\bf \Delta} = \hat{\bf \Delta}^{(0)} + \hat{\bf 
\Delta}^{(1)}$ ($\hat{\bf \Delta}^{(1)} \equiv \hat{\bf 
\Delta}^{(p)} + \hat{\bf \Delta}^{(t)}$). $\hat{\bf \Delta}^{(0)}$ 
is the leading part and the gradient part $\hat{\bf \Delta}^{(1)}$ 
represents variation in the macroscopic spacetime coordinates 
$X_\mu$, through first-order derivative $\partial_{X_\mu}$. 
Interactions among the fields give rise to reactions taking place in 
a system, which, in turn, causes a nontrivial change in the number 
density of quasiparticles. Thus, the self-energy part $\hat{\bf 
\Sigma}$ ties with $\hat{\bf \Delta}^{(1)}$. More precisely, 
$\hat{\bf \Sigma}^{- 1}$ is of the same order of magnitude as 
$\hat{\bf \Delta}^{(1)}$. Hence, in computing $\hat{\bf \Sigma}$ in 
the approximation under consideration, it is sufficient to keep the 
leading part (i.e., the part with no $X_\mu$-derivative): 
\begin{equation} 
\hat{\bf \Sigma} (x, y) \simeq {\bf P}_\xi (x) \hat{\Sigma}^{(\xi)} 
(x, y) {\bf P}_\xi (y) \simeq {\bf P}_\xi (X) \int \frac{d^{\, 4} 
P}{(2 \pi)^4} e^{- i P \cdot (x - y)} \hat{\Sigma}^{(\xi)} (X; P) 
\, . 
\label{sig-lea} 
\end{equation} 

We are adopting the loop-expansion. Then, to the gradient 
approximation, the relevant self-energy diagrams are the one-loop 
diagrams, together with relevant counter diagrams. There are two 
one-loop diagrams, the one is the tadpole diagram that includes one 
vertex coming from $\hat{\cal L}_{int}^{(4)}$ (cf. the definition 
(\ref{2.6d})), and the one is the diagram with two vertices 
coming from $\hat{\cal L}_{int}^{(3)}$. The counter diagrams are the 
diagrams that include $\hat{\cal L}_{r c} + \hat{\cal L}_{r c}$ (cf. 
Eq.~(\ref{owari}) with Eqs.~(\ref{L-q}) and (\ref{teigi-yo}).) The 
contribution from the tadpole diagram includes no 
$X_\mu$-derivative. Yet higher-order contributions to $\hat{\bf 
\Sigma}$ come in when one proceeds beyond the gradient 
approximation. 

A $\hat{\bf \Sigma}$-resummed propagator $\hat{\bf G}$ obeys the 
Schwinger-Dyson equation: 
\begin{equation} 
\hat{\bf G} = \hat{\bf \Delta} + \hat{\bf \Delta} 
\cdot \hat{\bf \Sigma} \cdot \hat{\bf G} \, . 
\label{SD} 
\end{equation} 
In Appendix B, we solve this within the gradient approximation. The 
result is 
\begin{eqnarray} 
\hat{\bf G} (x, y) & \simeq & \hat{\bf G}^{(0)} (x, y) + \int 
\frac{d^{\, 4} P}{(2 \pi)^4} e^{- i P \cdot (x - y)} \left[ \hat{\bf 
G}^{(p)} (X; P) + \hat{\bf G}^{(t)} (X; P) \right] \, , \nonumber \\ 
\hat{\bf G}^{(0)} (x, y) & = & {\bf P}_\xi (x) \int 
\frac{d^{\, 4} P}{(2 \pi)^4} \, 
e^{- i P \cdot (x - y)} \hat{G}^{(\xi)} (X; P) {\bf P}_\xi (y) \, , 
\nonumber \\ 
\hat{\bf G}^{(p)} (X; P) & = & \hat{\bf G}^{(p)}_1 (X; P) + 
\hat{\bf G}^{(p)}_2 (X; P) \, , \nonumber \\ 
\hat{\bf G}^{(p)}_1 (X; P) & = & - \frac{i}{2} {\bf P}_\xi (X) 
\left[ \left\{ f_\xi, \, P^2 - {\cal M}_\xi^2 - \Sigma_R^{(\xi)} 
\right\} \left( G_R^{(\xi)} \right)^2 + \left\{ f_\xi, \, P^2 - 
{\cal M}_\xi^2 - \Sigma_A^{(\xi)} \right\} \left( G_A^{(\xi)} 
\right)^2 \right] \hat{A}_+ \nonumber \\ 
\hat{\bf G}^{(p)}_2 (X; P) & = & - i {\bf P}_\xi (X) \left[ 
\tilde{\Gamma}_\xi^{(p)} - \left\{ f_\xi, \, P^2 - {\cal M}_\xi^2 - 
Re \Sigma_R^{(\xi)} \right\} \right] G_R^{(\xi)} G_A^{(\xi)} 
\hat{A}_+ \, , \nonumber \\ 
\hat{\bf G}^{(t)} (X; P) & = & 2 i | \hat{\varphi} \rangle P \cdot 
\stackrel{\leftrightarrow}{\partial} \langle \hat{\varphi}| 
\left( \begin{array}{cc} 
G_R^{(\pi)} G_R^{(\sigma)} & \;\; 0 \\ 
G_R^{(\pi)} G_R^{(\sigma)} - G_A^{(\pi)} G_A^{(\sigma)} & \;\; 
- G_A^{(\pi)} G_A^{(\sigma)} 
\end{array} 
\right) + 2 i \left[ |\hat{\varphi} \rangle P \cdot \partial \langle 
\hat{\varphi} | \omega_G - |\hat{\varphi} \rangle P \cdot 
\stackrel{\leftarrow}{\partial} \langle \hat{\varphi} | \omega_G' 
\right] \hat{A}_+ \, , 
\label{prop-sum} 
\end{eqnarray} 
where 
\begin{eqnarray} 
\hat{G}^{(\xi)} (X; P) & = & 
\left( 
\begin{array}{cc} 
G_R^{(\xi)} + f_\xi \left( G_R^{(\xi)} - G_A^{(\xi)} \right) & 
\;\;\; f_\xi \left( G_R^{(\xi)} - G_A^{(\xi)} \right) \\ 
\left( 1 + f_\xi \right) \left( G_R^{(\xi)} - G_A^{(\xi)} \right) & 
\;\;\; - G_A^{(\xi)} + f_\xi \left( G_R^{(\xi)} - G_A^{(\xi)} 
\right) 
\end{array} 
\right) \, , \nonumber \\ 
i \tilde{\Gamma}_\xi^{(p)} & = & (1 + f_\xi) \Sigma_{1 2}^{(\xi)} 
- f_\xi \Sigma_{2 1}^{(\xi)} \, , 
\label{6.44} 
\\ 
\omega_G & = & f_\pi G_R^{(\sigma)} \left( G_R^{(\pi)} - G_A^{(\pi)} 
\right) + f_\sigma G_A^{(\pi)} \left( G_R^{(\sigma)} - 
G_A^{(\sigma)} \right) \, , \nonumber \\ 
\omega_G' & = & \omega_G [\pi \leftrightarrow \sigma] \, . 
\label{prop-hoj} 
\end{eqnarray} 

\noindent 
In the above equations, $f_\xi = f_\xi (X; P)$, $\Sigma\mbox{'s} = 
\Sigma (X; P)$'s, ${\cal M}_\xi^2 = {\cal M}_\xi^2 (X)$, and 
\begin{eqnarray} 
G_{R (A)}^{(\xi)} & = & G_{R (A)}^{(\xi)} (X; P) = \frac{1}{P^2 - 
{\cal M}_\xi^2 (X) - \Sigma_{R (A)}^{(\xi)} (X; P)} \, , \nonumber 
\\ 
\Sigma_{R (A)}^{(\xi)} & = & \Sigma_{1 1}^{(\xi)} + \Sigma_{1 
2 (2 1)}^{(\xi)} \, . 
\label{R/A/y} 
\end{eqnarray} 
In Appendix C (cf. Eq.~(\ref{su1})), we show that $\Sigma_A^{(\xi)} 
(X; P) = \left( \Sigma_A^{(\xi)} (X; P) \right)^*$. The expression 
for $\Sigma_R^{(\xi)}$ is given in Appendix D, and $\Sigma_{1 
2}^{(\xi)}$ and $\Sigma_{2 1}^{(\xi)}$ in Eq.~(\ref{6.44}) on the 
mass-shell are computed in Appendix E, which plays a role in 
Sec.~VIC. 
\subsection{Gap equation} 
We have introduced two \lq\lq mass functions'' $M_\xi^2 (X)$ $(\xi = 
\pi, \sigma)$, which is a generalization of \cite{chiku}, in which 
only one mass parameter is introduced for computing an effective 
potential for the equilibrium system. In this subsection, we 
determine so far arbitrary mass functions. Various methods are 
available to this end (see, e.g., \cite{chiku} ). Among those we 
employ the on-shell renormalization scheme, 
\begin{equation} 
Re \Sigma_R^{(\xi)} (X; P) \, \rule[-3mm]{.14mm}{7mm} 
\raisebox{-2.85mm}{\scriptsize{$\; \mbox{s.p.}$}} 
\equiv Re \Sigma_R^{(\xi)} (X; p_0^2 = {\cal M}_\xi^2 (X), {\bf p} = 
{\bf 0}) = 0 
\label{alp} 
\end{equation} 
$(\xi = \pi, \sigma)$. For a subtraction point, \lq s.p.,' instead 
of $(p_0^2 = {\cal M}_\xi^2 (X), {\bf p} = {\bf 0})$ adopted here, 
we can choose $(p_0^2 = E_{p_s}^{(\xi) 2}, {\bf p} = {\bf p}_s)$, 
where $E_{p_s}^{(\xi)} = \sqrt{p^2_s +{\cal M}_\xi^2 (X)}$ with 
${\bf p}_s$ arbitrary. Eq.~(\ref{alp}) is the gap equation, by which 
$M_\xi^2 (X)$ $(\xi = \pi, \sigma)$ are determined self 
consistently. The explicit form of the gap equation (\ref{alp}) to 
leading one-loop order is displayed in Appendix D. $M_\xi^2 (X)$ 
thus determined depends on $\vec{\varphi}^{\, 2} (X)$ and $f_\xi$. 
With $M_\xi^2 (X)$ in hand, we can judge if ${\cal M}_\xi^2 (X)$, 
Eq.~(\ref{mass}), is positive or negative. If ${\cal M}_\xi^2 (X) < 
0$, the log-wave-length modes, $p^2 < |{\cal M}_\xi (X)^2|$, have 
imaginary frequencies and thus unstable, which causes perturbative 
instability of the system \cite{weinberg}. Unfortunately, no 
consistent scheme for treating such a case is available. Then, when 
${\cal M}_\xi^2 (X) < 0$ happens, we abandon the condition 
(\ref{alp}) and set ${\cal M}_\xi^2 (X) = 0$. Perturbation theory in 
this case is less efficient. 
\subsection{Boltzmann equation}
In order to find a physical meaning of $f_\xi$, we recall a 
momentum density of the system: 
\[ 
\vec{\cal P} (x) = - \mbox{Tr} \left[ \left\{ \frac{\partial 
\vec{\phi} (x)}{\partial x_0} \cdot \nabla \vec{\phi} (y) \right\} 
\rho \right]_{y = x} = - \frac{i}{2} \frac{\partial}{\partial x_0} 
\nabla_y G_c^{\alpha \alpha} (x, y) \, \rule[-3mm]{.14mm}{8.5mm} 
\raisebox{-2.85mm}{\scriptsize{$\; y = x$}} \, .  
\] 
We first analyze the contribution from ${\bf G}^{(0)}_c$ $(= {\bf 
G}^{(0)}_{1 1} + {\bf G}^{(0)}_{2 2})$, Eq.~(\ref{prop-sum}). When 
the interaction is switched off, 
${\bf G}^{(0)}_c$ reduces to ${\bf \Delta}^{(0)}_c$ $(= {\bf 
\Delta}^{(0)}_{1 1} + {\bf \Delta}^{(0)}_{2 2})$, 
Eq.~(\ref{prop-fin}). Computation of the contribution from ${\bf 
\Delta}^{(0)}_c$ yields 
\[ 
\vec{\cal P} (x) \, \rule[-3mm]{.14mm}{8.5mm} 
\raisebox{-2.85mm}{\scriptsize{$\; {\bf \Delta}^{(0)}_c$}} = 
\frac{c_\xi}{2} \int \frac{d^{\, 3} p}{(2 \pi)^3} \, \vec{p} \left\{ 
[f_\xi (x; E_p^{(\xi)}, \hat{\bf p}) + 1 / 2] + [f_\xi (x; - 
E_p^{(\xi)}, \hat{\bf p}) + 1 / 2] \right\} \, , 
\] 
where $c_\pi = N - 1$, $c_\sigma = 1$, $E_p^{(\xi)} = \sqrt{p^2 + 
{\cal M}_\xi^2}$, and $\hat{\bf p} = {\bf p} / |{\bf p}|$. 
Similarly, the computation of the contribution to the free ($\lambda 
= 0$) energy density yields 
\[ 
{\cal P}_{free}^0 (x) \, \rule[-3mm]{.14mm}{8.5mm} 
\raisebox{-2.85mm}{\scriptsize{$\; {\bf \Delta}^{(0)}_c$}} = 
\frac{c_\xi}{2} \int \frac{d^{\, 3} p}{(2 \pi)^3} \, E_p^{(\xi)} 
[f_\xi (x; E_p^{(\xi)}, \hat{\bf p}) - f_\xi (x; - E_p^{(\xi)}, 
\hat{\bf p}) ] \, . 
\] 
Subtracting the contribution from the vacuum, we see that $f_\xi$ 
$(\xi = \pi, \sigma)$ is related to the number density $N_\xi (x; 
E_p^{(\xi)}, {\bf p})$ of $\xi$ through 
\begin{equation} 
N_\xi (x, E_p^{(\xi)}, \hat{\bf p}) = f_\xi (x; E_p^{(\xi)}, 
\hat{\bf p}) = - 1 - f_\xi (x; - E_p^{(\xi)}, - \hat{\bf p}) \, . 
\label{kank} 
\end{equation} 
It is to be noted that the argument \lq $x$' here is macroscopic 
coordinates. For the interacting system, the corresponding relation 
is obtained using ${\bf G}_c$ $(= {\bf G}_{1 1} + {\bf G}_{2 2})$, 
Eq.~(\ref{prop-sum}), and the total energy density ${\cal P}^0 (x)$ 
in place of ${\cal P}^0_{free} (x)$ above. The contribution to 
${\cal P}^\mu (x)$ from the difference ${\bf G}^{(0)}_c - {\bf 
\Delta}^{(0)}_c$ yields a correction to the relation (\ref{kank}). 
This is also the case for the contribution of $({\bf G}_1^{(p)})_c$. 
(${\bf G}^{(t)})_c$ yields vanishing contribution to the gradient 
approximation. 

We now turn to analyzing the remaining contribution that comes 
from $({\bf G}_2^{(p)})_c$ in Eq.~(\ref{prop-sum}). Since $({\bf 
G}_2^{(p)})_c \propto G_R^{(\xi)} G_A^{(\xi)}$, in the narrow-width 
approximation, $Im \Sigma_R^{(\xi)} (X; P) \to - \epsilon (p_0) 
0^+$, pinch singularity is developed. Then, the contribution of 
$({\bf G}_2^{(p)})_c$ to ${\cal P}^\mu (x)$ diverges in this 
approximation. In practice, $Im \Sigma_R^{(\xi)}$ $(\propto 
\lambda^2)$ is a small quantity, so that the contribution, although 
not divergent, is large. This invalidates the perturbative scheme 
and a sort of \lq\lq renormalization'' is necessary for the number 
density \cite{nie}. This observation leads us to introduce a 
condition $({\bf G}_2^{(p)})_c = 0$ or 
\begin{equation} 
\left\{ f_\xi, \, P^2 - {\cal M}_\xi^2 - Re \Sigma_R^{(\xi)} 
\right\} = \tilde{\Gamma}_\xi^{(p)} \;\;\;\;\; (\xi = \pi, \sigma) 
\, . 
\label{yama} 
\end{equation} 
This serves as determining equation for so far arbitrary $f_\xi$. 
Then $\hat{\bf G}$ in Eq.~(\ref{prop-sum}) becomes $\hat{\bf G} = 
\hat{\bf G}^{(0)} + \hat{\bf G}^{(p)}_1 + \hat{\bf G}^{(t)}$, which 
is free from pinch singularity in the narrow-width approximation. 
It is obvious that, in the present scheme, above-mentioned large 
contributions do not appear. 

In order to disclose the physical meaning of Eq.~(\ref{yama}), we 
first define on the mass-shell, $p_0 = \pm \omega_\xi (X; \pm {\bf 
p})$ $(\equiv \pm \omega_\pm^{(\xi)})$: 
\begin{equation} 
Re \left( \Sigma_R^{(\xi)} (X; P) \right)^{- 1} \, 
\rule[-3mm]{.14mm}{8.5mm} \raisebox{-2.85mm}{\scriptsize{$\; p_0 = 
\pm \omega_\pm^{(\xi)}$}} = \left[ P^2 - {\cal M}_\xi^2 (X) - Re \, 
\Sigma_R^{(\xi)} (X; P) \right]_{p_0 = \pm \omega_\pm^{(\xi)}} = 0 
\, . 
\label{7.13} 
\end{equation} 
From Eq.~(\ref{alp}), we see that, when ${\cal M}_\xi^2 (X) > 0$, 
$\omega_\pm^{(\xi)} (X; {\bf 0}) = {\cal M}_\xi (X)$. We also 
introduce a wave-function renormalization factor, 
\[ 
Z^{- 1}_\xi \equiv 1 - \frac{1}{2 \omega_\xi (X; {\bf p})} 
\frac{\partial Re \Sigma_R^{(\xi)}}{\partial p_0} \, 
\rule[-3mm]{.14mm}{8.5mm} \raisebox{-2.85mm}{\scriptsize{$\; p_0 = 
\omega_\xi (X; {\bf p})$}} \, . 
\] 
It is now straightforward to show \cite{nie} that Eq.~(\ref{yama}) 
becomes, on the mass-shell, 
\begin{eqnarray} 
&& \frac{\partial N_\xi}{\partial X_0} 
+ {\bf v}_\xi \cdot \nabla_X N_\xi + 
\frac{\partial \omega_\xi (X; {\bf p})}{\partial X_\mu} 
\frac{\partial N_\xi}{\partial P^\mu} 
\, \rule[-3mm]{.14mm}{8.5mm} \raisebox{-2.85mm}{\scriptsize{$\; p_0 
= \omega_\xi (X; {\bf p})$}} \nonumber \\ 
&& \mbox{\hspace*{4ex}} = 
\frac{d N_\xi (X; \omega_\xi (X; {\bf p}), \hat{\bf p})}{d X_0} + 
\frac{\partial \omega_\xi (X; {\bf p})}{\partial {\bf p}} \cdot 
\frac{\partial N_\xi}{\partial {\bf X}} - 
\frac{\partial \omega_\xi (X; {\bf p})}{\partial {\bf X}} \cdot 
\frac{d N_\xi}{d {\bf p}} \nonumber \\ 
&& \mbox{\hspace*{4ex}} \simeq Z_\xi \, \Gamma^{(p)}_\xi 
\, \rule[-3mm]{.14mm}{8.5mm} \raisebox{-2.85mm}{\scriptsize{$\; p_0 
= \omega_\xi (X; {\bf p})$}} \, , 
\label{Bol} \\ 
&& \Gamma^{(p)}_\xi 
\, \rule[-3mm]{.14mm}{8.5mm} \raisebox{-2.85mm}{\scriptsize{$\; p_0 
= \omega_\xi (X; {\bf p})$}} = \frac{- i}{2 \omega_\xi (X; {\bf p})} 
\left[ ( 1 + N_\xi) \Sigma_{1 2}^{(\xi)} - N_\xi \Sigma_{2 
1}^{(\xi)} \right]  
\, \rule[-3mm]{.14mm}{8.5mm} \raisebox{-2.85mm}{\scriptsize{$\; p_0 
= \omega_\xi (X; {\bf p})$}} \, . 
\label{sei} 
\end{eqnarray} 

\noindent 
Here $N_\xi$ is as in Eq.~(\ref{kank}) with $E_p^{(\xi)} \to 
\omega_\xi (X; {\bf p})$ and ${\bf v}_\xi = \partial \omega_\xi (X; 
{\bf p}) / \partial {\bf p}$ is the velocity of the quasiparticle 
mode with momentum ${\bf p}$. $\Gamma_\xi^{(p)}$ in Eq.~(\ref{sei}) 
is the net production rate of the quasiparticle of momentum ${\bf 
p}$. In fact, $\Gamma_\xi^{(p)}$ is the difference between the 
production rate and the decay rate, so that $\Gamma_\xi^{(p)}$ is 
the net production rate. In the case of an equilibrium system, 
$\Gamma_\xi^{(p)} = 0$ (detailed balance formula). $N_\xi = N_\xi 
(X; \omega_\xi (X; {\bf p}), \hat{\bf p})$ here is essentially (the 
main part of) the relativistic Wigner function, and Eq.~(\ref{Bol}) 
is the generalized relativistic Boltzmann equation (cf. \cite{hei}). 

Let us suppose the case ${\cal M}_\xi^2 (X) >> Re 
\Sigma_R^{(\xi)} \, \rule[-2mm]{.14mm}{6mm} 
\raisebox{-1.85mm}{\scriptsize{$\; p_0 = \pm 
\omega_\pm^{({\xi})}$}}$. The solution to Eq.~(\ref{7.13}) is $p_0 
\simeq \pm E_p^{(\xi)}$. To one-loop order under consideration, only 
diagram that contributes to $\Sigma_{12 (21)}^{(\xi)}$ in 
Eq.~(\ref{sei}) is the one that includes two vertices coming from 
$\hat{\cal L}_{int}^{(3)}$. Thus, $\Sigma_{12 (21)}^{(\xi)}$ 
contains two $\Delta_{12 (21)}$'s, each of which contains on-shell 
$\delta$-function, $\delta (Q^2 - {\cal M}_\xi^2 (X))$ (cf. 
Eq.~(\ref{lead})). One can easily see then that 
$\Sigma_{1 2 (2 1)}^{(\xi)} \, \rule[-2mm]{.14mm}{5mm} 
\raisebox{-1.85mm}{\scriptsize{$\; p_0 \simeq E_p^{(\xi)}$}}$ 
vanishes unless ${\cal M}_\sigma > 2 {\cal M}_\pi$. $\Sigma_{1 2 (2 
1)}^{(\xi)} (X; P)$ at $p_0 = E_p^{(\xi)}$ $(\xi = \pi, \sigma)$ is 
computed in Appendix E. Let us turn to the case where ${\cal 
M}_\xi^2 (X) \leq O \Big( \Sigma_R^{(\xi)}  \, 
\rule[-2mm]{.14mm}{6mm} \raisebox{-1.85mm}{\scriptsize{$\; p_0 = \pm 
\omega_\pm^{(\xi)}$}} \Big)$ with $\xi = \pi$ or $\xi = \sigma$ or 
$\xi = \pi$ {\em and} $\xi = \sigma$. For a hard momentum, $p > O 
\Big( \Sigma_R^{(\xi)} \, \rule[-2mm]{.14mm}{6mm} 
\raisebox{-1.85mm}{\scriptsize{$\; p_0 = \pm \omega_\pm^{(\xi)}$}} 
\Big)$, the same statement as above holds. For computing $\Sigma_{1 
2 (2 1)}^{(\xi)} (X; P)$ with soft $P$, i.e., $|p_0|, \, p \leq O 
\Big( \Sigma_R^{(\xi)} \, \rule[-2mm]{.14mm}{6mm} 
\raisebox{-1.85mm}{\scriptsize{$\; p_0 = \pm \omega_\pm^{(\xi)}$}} 
\Big)$, above two $\Delta_{1 2 (2 1)}$'s contained in $\Sigma_{1 2 
(2 1)}^{(\xi)}$'s should be replaced by $G_{12 (21)}$'s 
\cite{bra-pis}. 

What we have shown here is that the requirement of the absence of 
$\hat{\bf G}_2^{(p)} \propto G_R^{(\xi)} G_A^{(\xi)}$ from $\hat{\bf 
G}$ leads to the Boltzmann equation for the 
quasiparticle-distribution functions. This means that the 
quasiparticles thus defined are the well-defined modes in the 
medium, in the sense that no large contribution appears in 
perturbation theory. Conversely, if we start with defining the 
quasiparticles such that their distribution functions subject to the 
Boltzmann equation, then, on the basis of them, well-defined 
perturbation theory may be constructed. 

Comparison of our derivation of the generalized Boltzmann equation 
(GBE) with those in related works is made in \cite{nie}. It is 
worth recapitulating here the comparison with the derivation in 
nonequilibrium thermo field dynamics (NETFD) \cite{ume}, which is a 
variant of nonequilibrium quantum field theory. We have imposed the 
condition (\ref{yama}) for determining $f_\xi$ or the number density 
$N_\xi$, so that pinch singularities (in narrow-width approximation) 
disappear. On the other hand, in NETFD, which employs the 
(space)time representation, the GBE is derived by imposing \lq\lq an 
on-shell renormalization condition'' for the propagator.  Since the 
pinch singularity is a singularity in momentum space, it is not 
immediately obvious how to translate this condition into the 
(space)time representation, as adopted in NETFD. Nevertheless, 
closer inspection of the structure of both formalisms tells us that 
our condition is in accord with the on-shell renormalization 
condition in NETFD. Incidentally, reconciliation of the NETFD with 
the $|p_0|$-prescription (cf. Appendix A), a notion in momentum 
space, remains as an open problem.  
\setcounter{equation}{0}
\setcounter{section}{6} 
\def\theequation{\mbox{\arabic{section}.\arabic{equation}}} 
\section{Conclusion and outlook} 
In this paper we have constructed from first principles a 
perturbative framework for computing reaction rates of the processes 
taking place in the $O (N)$ linear-sigma system in a broken phase. 
Only approximation we have employed is the so-called gradient 
approximation, so that the framework applies to the quasiuniform 
systems near equilibrium or the nonequilibrium quasistationary 
systems. 

The reactions taking place in the system causes a spacetime 
evolution of the system --- development of phase transition. This 
is the next subject following to the present analysis. At the final 
stage of such an analysis, one should check whether or not the 
rate of the phase-change of the system is too large, so that the 
gradient approximation adopted in this paper is violated.  
\section*{Acknowledgments}
This work was supported in part by a Grant-in-Aide for Scientific 
Research ((C)(2) (No.~12640287)) of the Ministry of Education, 
Science, Sports and Culture of Japan. 
\setcounter{equation}{0}
\setcounter{section}{1}
\section*{Appendix A: $|p_0|$-prescription} 
\def\theequation{\mbox{\Alph{section}\arabic{equation}}} 
One starts with $f_\xi (x, y)$ that is local in time, $f_\xi (x, y) 
= \delta (x_0 - y_0) \sum_{\tau = \pm} g^{(\tau)} ({\bf x}, {\bf y}; 
x_0)$. Here $\pm$ denotes positive/negative frequency part and the 
time coordinate \lq $x_0$' is of macroscopic. Then, in place of 
Eq.~(\ref{L-p}), we have, with obvious notation, 
\begin{eqnarray} 
{\bf L}' (x, y) & \simeq & i {\bf P}_\xi (X) \int \frac{d^{\, 3} 
p}{(2 \pi)^3} \, e^{i {\bf p} \cdot ({\bf x} - {\bf y})} \sum_{\tau 
= \pm} \left[ 2 i \delta' (x_0 - y_0) \partial_{X_0} g^{(\tau)}_\xi 
(X; {\bf p}) \right. \nonumber \\ 
&& \left. + \delta (x_0 - y_0) \left( 2 {\bf p} \cdot \partial_{\bf 
X} - \frac{\partial {\cal M}_\xi^2 (X)}{\partial {\bf X}} \cdot 
\partial_{\bf p} \right) g^{(\tau)}_\xi (X; {\bf p}) \right] 
\label{hara} 
\end{eqnarray} 

\noindent 
with $X_0 = x_0$. The form (\ref{hara}) is local in time. 
It is well known through the analyses of equilibrium case that the 
following replacement ($|p_0|$-prescription) should be made: 
\[ 
\theta (p_0) g^{(+)}_\xi (X; {\bf p}) 
+ \theta (- p_0) g^{(-)}_\xi (X; {\bf p}) \rightarrow f_\xi (X; P) 
\] 
with $f_\xi (X; P)$ as in (\ref{L-p}) in the text. Adopting this 
prescription, we attain Eq.~(\ref{L-p}). (See \cite{nie} for more 
details.) 
\setcounter{equation}{0}
\setcounter{section}{2}
\section*{Appendix B: Derivation of Eq.~(6.3)} 
\def\theequation{\mbox{\Alph{section}\arabic{equation}}} 
Here we solve the Schwinger-Dyson equation (\ref{SD}). 
Multiplying $\hat{\bf B}_L^{(0) - 1}$, Eq.~(\ref{B00}), from left 
and $\hat{\bf B}_R^{(0) - 1}$ from right, we obtain 
\begin{equation} 
\underline{\hat{\bf G}} = \underline{\hat{\bf \Delta}} + 
\underline{\hat{\bf \Delta}} \cdot \underline{\hat{\bf 
\Sigma}} \cdot \underline{\hat{\bf G}} \, , 
\label{SD1} 
\end{equation} 
where 
\[ 
\underline{\hat{\bf G}} \equiv \hat{\bf B}_L^{(0) - 1} \cdot 
\hat{\bf G} \cdot \hat{\bf B}_R^{(0) - 1} \, , \;\;\;\;\; 
\underline{\hat{\bf \Delta}} \equiv \hat{\bf B}_L^{(0) - 1} 
\cdot \hat{\bf \Delta} \cdot \hat{\bf B}_R^{(0) - 1} \, , 
\;\;\;\;\; \underline{\hat{\bf \Sigma}} \equiv \hat{\bf 
B}_R^{(0)} \cdot \hat{\bf \Sigma} \cdot \hat{\bf B}_L^{(0)} \, . 
\] 
Straightforward manipulation using Eqs.~(\ref{ato}) and 
(\ref{prop-fin}) yields 
\begin{eqnarray} 
\underline{\hat{\bf \Delta}} & = & \left( 
\begin{array}{cc} 
{\bf \Delta}_R & \;\; {\bf \Delta}_{\mbox{\scriptsize{off}}} \\ 
0 & \;\;  - {\bf \Delta}_A 
\end{array} 
\right) \, , 
\label{d-off} 
\\ 
{\bf \Delta}_{\mbox{\scriptsize{off}}} (x, y) & \simeq & 2 i | 
\hat{\varphi} (X) \rangle \partial_{X^\mu} \langle \hat{\varphi} (X) 
| \int \frac{d^{\, 4} P}{(2 \pi)^4} \, 
e^{- i P \cdot (x - y)} \, P^\mu \left( f_\sigma (X; P) - f_\pi (X; 
P) \right) \Delta_R^{(\sigma)} (X; P) \Delta_A^{(\pi)} (X; P) 
\nonumber \\ 
& & + 2 i | \hat{\varphi} (X) \rangle 
\stackrel{\leftarrow}{\partial}_{X^\mu} \langle \hat{\varphi} (X) | 
\int \frac{d^{\, 4} P}{(2 \pi)^4} \, e^{- i P \cdot (x - 
y)} \, P^\mu \left( f_\sigma (X; P) - f_\pi (X; P) \right) 
\Delta_R^{(\pi)} (X; P) \Delta_A^{(\sigma)} (X; P) \nonumber \\ 
&& + i {\bf P}_\xi (X) \int \frac{d^{\, 4} P}{(2 \pi)^4} \, e^{- i P 
\cdot (x - y)} \left\{ f_\xi, \, P^2 - {\cal M}_\xi^2 \right\} 
\Delta_R^{(\xi)} (X; P) \Delta_A^{(\xi)} (X; P) \, . \nonumber 
\end{eqnarray} 
Using the identity (\ref{self-rel}) in Appendix C, we obtain 
\begin{eqnarray} 
\underline{\hat{\bf \Sigma}} & = & \left( 
\begin{array}{cc} 
{\bf \Sigma}_R & \;\;\; {\bf \Sigma}_{\mbox{\scriptsize{off}}} \\ 
0 & \;\;\; - {\bf \Sigma}_A 
\end{array}
\right) \, , 
\label{sigma-u} 
\\ 
{\bf \Sigma}_R & = & {\bf \Sigma}_{1 1} + {\bf \Sigma}_{1 2} \, , 
\;\;\;\;\; {\bf \Sigma}_A = {\bf \Sigma}_{1 1} + {\bf \Sigma}_{2 1} 
= - {\bf \Sigma}_{2 2} - {\bf \Sigma}_{1 2} \, , 
\label{SA} 
\\ 
{\bf \Sigma}_{\mbox{\scriptsize{off}}} & = & {\bf \Sigma}_{1 2} 
\cdot (1 + {\bf P}_\xi \cdot f_\xi \cdot {\bf P}_\xi) - {\bf P}_\xi 
\cdot f_\xi \cdot {\bf P}_\xi \cdot {\bf \Sigma}_{2 1} + {\bf 
\Sigma}_{1 1} \cdot {\bf P}_\xi \cdot f_\xi \cdot {\bf P}_\xi - {\bf 
P}_\xi \cdot f_\xi \cdot {\bf P}_\xi \cdot {\bf \Sigma}_{1 1} \, . 
\nonumber 
\end{eqnarray} 
Substituting the leading-order expression (\ref{sig-lea}) for 
${\bf \Sigma}$'s and using Eq.~(\ref{self-rel}) in Appendix C, we 
obtain, to the gradient approximation, 
\begin{eqnarray*} 
{\bf \Sigma}_{\mbox{\scriptsize{off}}} & \simeq & {\bf P}_\xi (x) 
\int \frac{d^{\, 4} P}{(2 \pi)^4} \, e^{- i P \cdot (x - y)} \left[ 
(1 + f_\xi (X; P)) \Sigma_{1 2}^{(\xi)} (X; P) - f_\xi (X; P) 
\Sigma_{2 1}^{(\xi)} (X; P) \right] {\bf P}_\xi (y) \\ 
&& + \frac{i}{2} {\bf P}_\xi (x) \int \frac{d^{\, 4} P}{(2 \pi)^4} 
\, e^{- i P \cdot (x - y)} \left\{ f_\xi , \, \Sigma_{1 1}^{(\xi)} - 
\Sigma_{2 2}^{(\xi)} \right\} {\bf P}_\xi (y) \, , 
\end{eqnarray*} 

\noindent 
where use has been made of ${\bf P}_\xi (\partial {\bf P}_\xi) {\bf 
P}_\xi = 0$ (no summation over $\xi$). To leading order under 
consideration, $\Sigma_{1 2}^{(\xi)} (X; P)$ and $\Sigma_{2 
1}^{(\xi)} (X; P)$ are pure imaginary. Then, from 
Eq.~(\ref{self-rel}), $Re \Sigma_{2 2}^{(\xi)} (X; P) = - Re 
\Sigma_{1 1}^{(\xi)} (X; P)$ and, from Eq.~(\ref{su1}) and 
Eq.~(\ref{SA}), we obtain $Im \Sigma_{2 2}^{(\xi)} (X; P) = Im 
\Sigma_{1 1}^{(\xi)} (X; P) = i (\Sigma_{1 2}^{(\xi)} (X; P) + 
\Sigma_{2 1}^{(\xi)} (X; P)) / 2$. Using these relations, we finally 
obtain 
\begin{eqnarray} 
&& {\bf \Sigma}_{\mbox{\scriptsize{off}}} (x, y) \simeq {\bf P}_\xi 
(x) \int \frac{d^{\, 4} P}{(2 \pi)^4} \, e^{- i P \cdot (x - y)} \, 
\Sigma_{\mbox{\scriptsize{off}}}^{(\xi)} (X; P) {\bf P}_\xi (y) \, , 
\nonumber \\ 
&& \Sigma_{\mbox{\scriptsize{off}}}^{(\xi)} (X; P) \simeq i 
\tilde{\Gamma}_\xi^{(p)} (X; P) + i \left\{ f_\xi , \, Re 
\Sigma_R^{(\xi)} \right\} \, . 
\label{pro} 
\end{eqnarray} 
Note that the first term of the R.H.S. of Eq.~(\ref{pro}), being 
proportional to $\lambda^2$, is proportional to the net-production 
rate (cf. above after Eq.~(\ref{sei})), which causes the change in 
the number density. The second term is proportional to $\lambda$ and 
includes derivatives with respect to $X_\mu$, and is of higher 
order. Then one can drop the second term. Nevertheless, we shall 
keep it in the following. 

Substituting Eqs.~(\ref{d-off}) and (\ref{sigma-u}) into 
Eq.~(\ref{SD1}) and retaining up to the terms that are linear in 
${\bf \Sigma}_{\mbox{\scriptsize{off}}}$, we obtain 
\begin{eqnarray} 
\underline{\hat{\bf G}} & = & \underline{\hat{\bf \Delta}} + \sum_{n 
= 1}^\infty \underline{\hat{\bf \Delta}} \left[ \cdot 
\underline{\hat{\bf \Sigma}} \cdot \underline{\hat{\bf \Delta}} 
\right]^n \nonumber \\ 
& \simeq & \underline{\hat{\bf \Delta}} + \sum_{n = 1}^\infty 
\underline{\hat{\bf \Delta}} \left[ \cdot \left( 
\begin{array}{cc} 
{\bf \Sigma}_R & \;\; 0 \\ 
0 & \;\; - {\bf \Sigma}_A 
\end{array} 
\right) 
\cdot \underline{\hat{\bf \Delta}} \right]^n 
+ \left( 
\begin{array}{cc} 
{\bf G}_R & \;\; 0 \\ 
0 & \;\; - {\bf G}_A 
\end{array} 
\right) 
\cdot \left( 
\begin{array}{cc} 
0 & \;\; {\bf \Sigma}_{\mbox{\scriptsize{off}}} \\ 
0 & \;\; 0 
\end{array} 
\right) \cdot 
\left( 
\begin{array}{cc} 
{\bf G}_R & \;\; 0 \\ 
0 & \;\; - {\bf G}_A 
\end{array} 
\right) \, , 
\label{SDD} 
\end{eqnarray} 
where higher-order terms have been dropped and 
\[ 
{\bf G}_{R (A)} = {\bf \Delta}_{R (A)} + {\bf \Delta}_{R (A)} 
\sum_{n = 1}^\infty \left( \cdot {\bf \Sigma}_{R (A)} \cdot {\bf 
\Delta}_{R (A)} \right)^n \, . 
\] 
Keeping terms linear in ${\bf \Delta}_{off}$ (Eq.~(\ref{d-off})), we 
can solve Eq.~(\ref{SDD}): 
\begin{eqnarray*} 
\left( \underline{\bf G} (x, y) \right)_{2 1} & = & 0 \, , \\ 
\left( \underline{\bf G} (x, y) \right)_{1 1} & = & {\bf P}_\xi (x) 
\int \frac{d^{\, 4} P}{(2 \pi)^4} \, e^{- i P \cdot (x - y)} \, 
G_R^{(\xi)} (X; P) {\bf P}_\xi (y) \\ 
&& + 2 i | \hat{\varphi} (X) \rangle 
\stackrel{\leftrightarrow}{\partial}_{X^\mu} \langle \hat{\varphi} 
(X) | \int \frac{d^{\, 4} P}{(2 \pi)^4} \, e^{- i P \cdot (x - 
y)} \, P^\mu G_R^{(\pi)} (X; P) G_R^{(\sigma)} (X; P) \, , \\ 
\left( \underline{\bf G} (x, y) \right)_{2 2} & = & - 
\left( \underline{\bf G} (x, y) \right)_{1 1} \, 
\rule[-3mm]{.14mm}{8.5mm} \raisebox{-2.85mm}{\scriptsize{$\; R \to 
A$}} \, , \\ 
\left( \underline{\bf G} (x, y) \right)_{1 2} & = & 2 i | 
\hat{\varphi} (X) \rangle \partial_{X^\mu} \langle \hat{\varphi} (X) 
| \int \frac{d^{\, 4} P}{(2 \pi)^4} \, e^{- i P \cdot (x - 
y)} \, P^\mu \left( f_\sigma (X; P) - f_\pi (X; P) \right) 
G_R^{(\sigma)} (X; P) G_A^{(\pi)} (X; P) \nonumber \\ 
&& + 2 i | \hat{\varphi} (X) \rangle 
\stackrel{\leftarrow}{\partial}_{X^\mu} \langle \hat{\varphi} (X) | 
\int \frac{d^{\, 4} P}{(2 \pi)^4} \, e^{- i P \cdot (x - 
y)} \, P^\mu \left( f_\sigma (X; P) - f_\pi (X; P) \right) 
G_R^{(\pi)} (X; P) G_A^{(\sigma)} (X; P) \\ 
&& - i {\bf P}_\xi (X) \int \frac{d^{\, 4} P}{(2 \pi)^4} \, e^{- i P 
\cdot (x - y)} \left[ \tilde{\Gamma}_\xi^{(p)} - \left\{ f_\xi , \, 
Re \left( G_R^{(\xi)} \right)^{- 1} \right\} \right] G_R^{(\xi)} (X; 
P) G_A^{(\xi)} (X; P) \, , 
\end{eqnarray*} 

\noindent 
where $G_{R (A)}^{(\xi)} (X; P)$ is as in Eq.~(\ref{R/A/y}) in the 
text. Computing $\hat{\bf G} = \hat{\bf B}_L^{(0)} \cdot 
\underline{\hat{\bf G}} \cdot \hat{\bf B}_R^{(0)}$ in the gradient 
approximation, we obtain Eqs.~(\ref{prop-sum})~-~(\ref{prop-hoj}) in 
the text. 
\setcounter{equation}{0}
\setcounter{section}{3}
\section*{Appendix C: Properties of the self-energy part} 
\def\theequation{\mbox{\Alph{section}\arabic{equation}}} 
It is obvious that the relation (\ref{wa}) holds for the full 
propagator $\hat{\bf G}$ and also for the self-energy-part inserted 
propagator $\hat{\bf \Delta} \cdot \hat{\bf \Sigma} \cdot \hat{\bf 
\Delta}$. From the latter, one can readily obtain the relation: 
\begin{equation} 
\sum_{i, \, j = 1}^2 {\bf \Sigma}_{i j} = 0 \, . 
\label{self-rel} 
\end{equation} 
The expression for full ${\bf G}_R$ and ${\bf G}_A$ with 
$\vec{\phi}_1 = \vec{\phi}_2$ $(\equiv \vec{\phi})$ are given by 
Eq.~(\ref{doku1}) with Heisenberg fields for $\phi$'s. From 
Eq.~(\ref{doku1}), we obtain 
\begin{equation} 
\left( \Delta_{R (A)}^{\alpha \beta} (x, y) \right)^* = 
\Delta_{R (A)}^{\alpha \beta} (x, y) \, , \;\;\;\;\;\;\;\; 
\Delta_A^{\beta \alpha} (y, x) = \Delta_R^{\alpha \beta} (x, y) 
\, . 
\label{ku} 
\end{equation} 
Eq.~(\ref{ku}) is also valid for self-energy-part-inserted 
propagator, $\hat{\bf \Delta} \cdot \hat{\bf \Sigma} \cdot 
\hat{\bf \Delta}$: 
\begin{eqnarray*} 
&& \left( \Delta_{R (A)}^{\alpha \alpha'} \cdot \Sigma_{R 
(A)}^{\alpha' \beta'} \cdot \Delta_{R (A)}^{\beta' \beta} \right)^* 
= \Delta_{R (A)}^{\alpha \alpha'} \cdot \Sigma_{R (A)}^{\alpha' 
\beta'} \cdot \Delta_{R (A)}^{\beta' \beta} \, , \\ 
&& \left[ \Delta_A^{\beta \beta'} \cdot  \Sigma_A^{\beta' \alpha'} 
\cdot \Delta_A^{\alpha' \alpha} \right] (y, x) = \left[ 
\Delta_R^{\alpha \alpha'} \cdot  \Sigma_R^{\alpha' \beta'} \cdot 
\Delta_R^{\beta' \beta} \right] (x, y) \, , 
\end{eqnarray*} 
which yields 
\begin{equation} 
\left( \left[ \Delta_R^{\alpha \alpha'} \cdot \Sigma_R^{\alpha' 
\beta'} \cdot \Delta_R^{\beta' \beta} \right] (x, y) \right)^* 
= \left[ \Delta_A^{\beta \beta'} \cdot 
\Sigma_A^{\beta' \alpha'} \cdot \Delta_A^{\alpha' \alpha} \right] 
(y, x) \, . 
\label{ke1} 
\end{equation} 
Applying $\partial^2 + {\cal M}_\xi^2 {\bf P}_\xi$ from both sides 
of Eq.~(\ref{ke1}), we obtain 
\begin{equation} 
\left( \Sigma_R^{\alpha \beta} (x, y) \right)^* = \Sigma_A^{\beta 
\alpha} (x, y) \, . 
\label{su} 
\end{equation} 
As discussed in Sec.~VIA, it is sufficient to compute the leading 
part of ${\bf \Sigma}_{i j}$, Eq.~(\ref{sig-lea}). Then, from 
Eq.~(\ref{su}), we obtain 
\begin{equation} 
\Sigma_A^{(\xi)} (X; P) = \left( \Sigma_R^{(\xi)} (X; P) \right)^* 
\, . 
\label{su1} 
\end{equation} 
\setcounter{equation}{0}
\setcounter{section}{4}
\section*{Appendix D: One-loop formula for ${\bf \Sigma}_R$ and the 
gap equation} 
\def\theequation{\mbox{\Alph{section}\arabic{equation}}} 
\subsection{Form for ${\bf \Sigma}_R (X; P)$} 
Here we display the concrete form of ${\bf \Sigma}_R (X; P)$ to 
one-loop order. For the relevant diagrams, see above after 
Eq.~(\ref{sig-lea}). For a UV-renormalization scheme, as in 
\cite{chiku}, we use the $\overline{\mbox{MS}}$ scheme. Computation 
is a straightforward generalization of \cite{chiku} and the final 
result reads 
\begin{eqnarray} 
{\bf \Sigma}_R (X; P) & = & {\bf P}_\xi (X) \Sigma_R^{(\xi)} (X; P) 
- \chi_\xi (X) \, , 
\label{kaii}  \\ 
\Sigma_R^{(\pi)} (X; P) & = & \frac{\lambda}{6} \left[ (N + 1) {\cal 
I}_\pi + {\cal I}_\sigma \right] + \frac{\lambda^2}{9} \varphi^2 (X) 
{\cal J}_{\pi \sigma} (X; P) \, , \nonumber \\ 
\Sigma_R^{(\sigma)} (X; P) & = & \frac{\lambda}{6} \left[ (N - 1) 
{\cal I}_\pi + 3 {\cal I}_\sigma \right] 
+ \frac{\lambda^2}{18} \varphi^2 (X) \left[ (N - 1) {\cal J}_{\pi 
\pi} (X; P) + 9 {\cal J}_{\sigma \sigma} (X; P) \right] \, . 
\nonumber 
\end{eqnarray} 
Here 
\begin{eqnarray} 
{\cal I}_\xi (X) & = & \frac{{\cal M}_\xi^2 (X)}{(4 \pi)^2} \, \ln 
\frac{{\cal M}_\xi^2 (X)}{e \mu_d^2} + \int \frac{d^{\, 3} q}{(2 
\pi)^3} \, \frac{N_\xi (X; E_q^{(\xi)} (X), \hat{\bf 
q})}{E_q^{(\xi)} (X)} \;\;\;\;\; (\xi = \pi, \sigma) \, , \nonumber 
\\ 
{\cal J}_{\pi \sigma} (X; P) & = & \frac{1}{(4 \pi)^2} \left[ 
\ln \frac{{\cal M}_\sigma^2 (X)}{e \mu_d^2} - \frac{P^2 - {\cal 
M}_\sigma^2 (X) + {\cal M}_\pi^2 (X)}{2 P^2} \, \ln 
\frac{{\cal M}_\sigma^2 (X)}{{\cal M}_\pi^2 (X)} \right. 
\nonumber \\ 
& & \left. - \frac{{\cal S}}{2 P^2} \ln \frac{\left( {\cal S} - P^2 
\right)^2 - \left( {\cal M}_\sigma^2 (X) - {\cal M}_\pi^2 (X) 
\right)^2}{\left( {\cal S} + P^2 \right)^2 - \left( {\cal 
M}_\sigma^2 (X) - {\cal M}_\pi^2 (X) \right)^2} \right] \nonumber \\ 
&& + \frac{1}{2} \int \frac{d^{\, 3} q}{(2 \pi)^3} \left\{ 
\frac{1}{E_q^{(\sigma)}} \left[ \frac{N_\sigma (X; E_q^{(\sigma)}, 
\hat{\bf q})}{P^2 + {\cal M}_\sigma^2 - {\cal M}_\pi^2 - 2 p_0 
E_q^{(\sigma)} + 2 {\bf p} \cdot {\bf q} + i \epsilon (p_0 - 
E_q^{(\sigma)}) 0^+} \right. \right. \nonumber \\ 
&& \left. \left. + \frac{N_\sigma (X; E_q^{(\sigma)}, \hat{\bf 
q})}{P^2 + {\cal M}_\sigma^2 - {\cal M}_\pi^2 + 2 p_0 E_q^{(\sigma)} 
- 2 {\bf p} \cdot {\bf q} + i \epsilon (p_0 + E_q^{(\sigma)}) 0^+} 
\right] + \left( \sigma \leftrightarrow \pi \right) \right\} \, , 
\label{jei} \\ 
{\cal J}_{\xi \xi} (X; P) & = & {\cal J}_{\pi \sigma} (X; P) 
\, \rule[-3mm]{.14mm}{7.5mm} \raisebox{-2.85mm}{\scriptsize{$\; 
{\cal M}_\pi^2 = {\cal M}_\sigma^2 \to {\cal M}_\xi^2$}} 
\;\;\;\;\;\;\;\;\; (\xi = \pi, \sigma) \, , 
\end{eqnarray} 
where $E_q^{(\xi)} (X) = \sqrt{q^2 + {\cal M}_\xi^2 (X)}$, 
\[ 
{\cal S} = \sqrt{ \left[ \left\{ {\cal M}_\sigma (X) - {\cal M}_\pi 
(X) \right\}^2 - P^2 \right] \left[ \left\{ {\cal M}_\sigma (X) + 
{\cal M}_\pi (X) \right\}^2 - P^2 \right]} \, , 
\] 
and $\mu_d$ is an arbitrary parameter that appears in the 
dimensional-regularization scheme, as adopted here. The first term 
on the R.H.S. of Eq.~(\ref{jei}) is valid in the region $P^2 < 0$. 
Its expressions in other regions of $P^2$ are obtained through 
analytic continuation with ${\cal M}_\xi^2 (X) \to {\cal M}_\xi^2 
(X) - i \epsilon (p_0) 0^+$. 
\subsection{Gap equation} 
The gap equation (\ref{alp}) with Eq.~(\ref{kaii}) yields 
\begin{equation} 
M_\xi^2 (X) - m^2 = \delta M_\xi^2 (X) \;\;\;\;\;\; (\xi = \pi, \, 
\sigma) 
\label{cond-22} 
\end{equation} 
with 
\begin{eqnarray} 
\delta M_\pi^2 (X) & = & \frac{\lambda}{6} \left[ (N + 1) {\cal 
I}_\pi + {\cal I}_\sigma \right] + \frac{\lambda^2}{9} \, \varphi^2 
(X) \left[ \frac{1}{(4 \pi)^2} \left\{ \frac{{\cal M}_\sigma^2 
(X)}{2 {\cal M}_\pi^2 (X)} \, \ln \frac{{\cal M}_\sigma^2 (X)}{{\cal 
M}_\pi^2 (X)} + \ln \frac{{\cal M}_\pi^2 (X)}{e^2 \mu_d^2} - {\cal 
K}_{\pi \sigma}^{(\pi)} \right\} \right. \nonumber \\ 
& & \left. + {\cal H}_{\pi \sigma}^{(\beta)} (X; p_0 = {\cal M}_\pi 
(X), {\bf 0}) \right] \, , 
\label{ai-1} \\ 
\delta M_\sigma^2 (X) & = & \frac{\lambda}{6} \left[ (N - 1) 
{\cal I}_\pi + 3 {\cal I}_\sigma \right] + \frac{ \lambda^2}{18} \, 
\varphi^2 (X) \left[ \frac{1}{(4 \pi)^2} \left\{ (N - 1) \left( \ln 
\frac{ {\cal M}_\pi^2 (X)}{e^2 \mu_d^2 } - {\cal K}_{\pi 
\pi}^{(\sigma)} \right) + 3 \sqrt{3} \pi + 9 \, \ln \frac{{\cal 
M}_\sigma^2 (X)}{e^2 \mu_d^2} \right\} \right. \nonumber \\ 
& & \left. + \left\{ (N - 1) {\cal H}_{\pi \pi}^{(\beta)} (X; p_0 
= {\cal M}_\sigma (X), {\bf 0}) + 9 {\cal H}_{\sigma 
\sigma}^{(\beta)} (X; p_0 = {\cal M}_\sigma (X), {\bf 0}) \right\} 
\right] \, . \nonumber \\ 
\label{ai-2} 
\end{eqnarray} 
Here 
\begin{eqnarray*} 
{\cal K}_{\pi \sigma}^{(\pi)} & = & \theta \left( {\cal M}_\sigma^2 
(X) - 4 {\cal M}_\pi^2 (X) \right) \, \frac{\sqrt{{\cal M}_\sigma^2 
(X) ({\cal M}_\sigma^2 (X) - 4 {\cal M}_\pi^2 (X))}}{2 \, {\cal 
M}_\pi^2 (X)} \\ 
& & \times \left[ \ln \frac{{\cal M}_\sigma^2 (X) - \sqrt{{\cal 
M}_\sigma^2 (X) ({\cal M}_\sigma^2 (X) - 4 {\cal M}_\pi^2 
(X))}}{{\cal M}_\sigma^2 (X) + \sqrt{{\cal M}_\sigma^2 (X) 
({\cal M}_\sigma^2 (X) - 4 {\cal M}_\pi^2 (X))}} \right. \\ 
& & \left. - \ln \frac{{\cal M}_\sigma^2 (X) - 2 \, {\cal M}_\pi^2 
(X) - \sqrt{{\cal M}_\sigma^2 (X) ( {\cal M}_\sigma^2 (X) - 4 {\cal 
M}_\pi^2 (X))}}{{\cal M}_\sigma^2 (X) - 2 \, {\cal M}_\pi^2 (X) + 
\sqrt{{\cal M}_\sigma^2 (X) ( {\cal M}_\sigma^2 (X) - 4 {\cal 
M}_\pi^2 (X))}} \right] \, , \\ 
& & + \theta \left( 4 {\cal M}_\pi^2 (X) - {\cal M}_\sigma^2 (X) 
\right) \, \frac{\sqrt{{\cal M}_\sigma^2 (X) (4 {\cal M}_\pi^2 (X) - 
{\cal M}_\sigma^2 (X))}}{{\cal M}_\pi^2 (X)} \\ 
& & \times \left[ \epsilon ({\cal M}_\sigma^2 (X) ) \arctan 
\frac{\sqrt{{\cal M}_\sigma^2 (X) (4 {\cal M}_\pi^2 (X) - {\cal 
M}_\sigma^2 (X))}}{{\cal M}_\sigma^2 (X)} \right. \\ 
& & - \epsilon ({\cal M}_\sigma^2 (X) - 2 {\cal M}_\pi^2 (X) 
) \arctan \frac{\sqrt{{\cal M}_\sigma^2 (X) (4 {\cal M}_\pi^2 (X) - 
{\cal M}_\sigma^2 (X))}}{{\cal M}_\sigma^2 (X) - 2 {\cal M}_\pi^2 
(X)} \\ 
& & \left. - \pi \, \theta \left( 2 {\cal M}_\pi^2 (X) - {\cal 
M}_\sigma^2 (X) \right) \right] \, , 
\\ 
{\cal K}_{\pi \pi}^{(\sigma)} & = & \theta \left( {\cal M}_\sigma^2 
(X) - 4 {\cal M}_\pi^2 (X) \right) \\ 
& & \times \frac{\sqrt{{\cal M}_\sigma^2 (X) ({\cal M}_\sigma^2 (X) 
- 4 {\cal M}_\pi^2 (X))}}{{\cal M}_\sigma^2 (X)} \, \ln \frac{{\cal 
M}_\sigma^2 (X) - \sqrt{{\cal M}_\sigma^2 (X) ({\cal M}_\sigma^2 (X) 
- 4 {\cal M}_\pi^2 (X))}}{{\cal M}_\sigma^2 (X) + \sqrt{{\cal 
M}_\sigma^2 (X) ({\cal M}_\sigma^2 (X) - 4 {\cal M}_\pi^2 (X))}} \\ 
& & + \theta \left( 4 {\cal M}_\pi^2 (X) - {\cal M}_\sigma^2 (X) 
\right) \, \frac{\sqrt{{\cal M}_\sigma^2 (X) (4 {\cal M}_\pi^2 (X) - 
{\cal M}_\sigma^2 (X))}}{{\cal M}_\sigma^2 (X)} \\ 
& & \times \left[ 2 \arctan \frac{\sqrt{{\cal M}_\sigma^2 (X) (4 
{\cal M}_\pi^2 (X) - {\cal M}_\sigma^2 (X))}}{{\cal M}_\sigma^2 
(X)} - \pi \right] \, , 
\end{eqnarray*} 
and 
\begin{eqnarray*} 
{\cal H}_{\pi \sigma}^{(\beta)} (X; p_0, {\bf 0}) & = & \int 
\frac{d^{\, 3} q}{(2 \pi)^3} \left[ \frac{{\cal M}_\sigma^2 (X) \, 
N_\sigma (X; E_q^{(\sigma)} (X), \hat{\bf q})}{E_q^{(\sigma)} (X) 
\left\{ {\cal M}_\sigma^4 (X) - 4 p_0^2 (E_q^{(\sigma)} (X))^2 
\right\}} \right. \\ 
&& \left. + \frac{(2 {\cal M}_\pi^2 (X) - {\cal M}_\sigma^2 (X)) \, 
N_\pi (X; E_q^{(\pi)} (X), \hat{\bf q})}{E_q^{(\pi)} (X) \left\{ 
(2{\cal M}_\pi^2 (X) - {\cal M}_\sigma^2 (X))^2 - 4 p_0^2 
(E_q^{(\pi)} (X))^2 \right\}} \right] \, , \\ 
{\cal H}_{\xi \xi}^{(\beta)} (X; p_0, {\bf 0}) & = & 2 \int 
\frac{d^{\, 3} q}{(2 \pi)^3} \frac{{\cal M}_\xi^2 (X) \, N_\xi (X; 
E_q^{(\xi)} (X), \hat{\bf q})}{E_q^{(\xi)} (X) \left\{ {\cal 
M}_\xi^4 (X) - 4 p_0^2 (E_q^{(\xi)} (X))^2 \right\}} \;\;\;\;\; 
(\xi = \pi, \sigma) \, . 
\end{eqnarray*} 
\setcounter{equation}{0}
\setcounter{section}{5}
\section*{Appendix E: Computation of ${\bf \Sigma}_{1 2}$ and ${\bf 
\Sigma}_{2 1}$} 
\def\theequation{\mbox{\Alph{section}\arabic{equation}}} 
Here, we compute the leading part of the one-loop contribution to 
${\bf \Sigma}_{1 2 (2 1)}$: 
\begin{eqnarray*} 
{\bf \Sigma}_{1 2 (2 1)} (X; P) & = & {\bf P}_\pi (X) \Sigma_{1 2 (2 
1)}^{(\pi)} (X; P)  + {\bf P}_\sigma (X) \Sigma_{1 2 (2 
1)}^{(\sigma)} (X; P)  \, , \\ 
\Sigma_{1 2 (2 1)}^{(\pi)} (X; P) & = & - \frac{i \lambda^2 
\varphi^2 (X)}{9} \int \frac{d^{\, 4} Q}{(2 \pi)^4} \, \Delta_{1 2 
(2 1)}^{(\sigma)} (X; Q) \Delta_{1 2 (2 1)}^{(\pi)} (X; P - Q) \, , 
\\ 
\Sigma_{1 2 (2 1)}^{(\sigma)} (X; P) & = & - \frac{i \lambda^2 
\varphi^2 (X)}{1 8} \int \frac{d^{\, 4} Q}{(2 \pi)^4} \left[ (N - 1) 
\Delta_{1 2 (2 1)}^{(\pi)} (X; Q) \Delta_{1 2 (2 1)}^{(\pi)} (X; P - 
Q) \right. \\ 
&& \left. + 9 \Delta_{1 2 (2 1)}^{(\sigma)} (X; Q) \Delta_{1 2 (2 
1)}^{(\sigma)} (X; P - Q) \right] \, . 
\end{eqnarray*} 
We compute $\Sigma_{1 2 (2 1)}^{(\xi)}$ on the mass-shell $p_0 = 
E_p^{(\xi)}$. The contribution to $\Sigma_{1 2 (2 1)}^{(\sigma)}$ 
from the term that accompanies two $\Delta_{1 2 (2 1)}^{(\sigma)}$'s 
vanishes. Nonvanishing contributions emerge only when $2 {\cal 
M}_\pi < {\cal M}_\sigma$. Computation is straightforward but 
lengthy. We only display the final forms: 
\begin{eqnarray} 
\Sigma_{1 2}^{(\pi)} (X; P) & = & \frac{i \lambda^2}{7 2 \pi} 
\frac{\varphi^2 (X)}{p} \int_{\xi_{l1}}^{\xi_{u1}} d \xi \, N_\sigma 
(X; \xi, \hat{\bf q}) \left[ 1 + N_\pi (X; \xi - E_p^{(\pi)}, 
\widehat{{\bf q} - {\bf p}}) \right]_{\hat{\bf p} \cdot 
\hat{\bf q} = z_1} \, , \nonumber \\ 
\Sigma_{2 1}^{(\pi)} (X; P) & = & \frac{i \lambda^2}{7 2 \pi} 
\frac{\varphi^2 (X)}{p} \int_{\xi_{l2}}^{\xi_{u2}} d \xi \, N_\pi 
(X; \xi, \hat{\bf q}) \left[ 1 + N_\sigma (X; \xi + E_p^{(\pi)}, 
\widehat{{\bf q} + {\bf p}}) \right]_{\hat{\bf p} \cdot 
\hat{\bf q} = z_2} \, , \nonumber \\ 
\Sigma_{1 2}^{(\sigma)} (X; P) & = & \frac{i (N - 1) \lambda^2}{1 4 
4 \pi} \frac{\varphi^2 (X)}{p} \int_{\xi_{l3}}^{\xi_{u3}} d \xi \, 
N_\pi (X; \xi, \hat{\bf q}) N_\pi (X; E_p^{(\sigma)} - \xi, 
\widehat{{\bf p} - {\bf q}}) \, \rule[-3mm]{.14mm}{7.5mm} 
\raisebox{-2.85mm}{\scriptsize{$\; \hat{\bf p} \cdot \hat{\bf q} = 
z_3$}} \, , \nonumber \\ 
\Sigma_{2 1}^{(\sigma)} (X; P) & = & \frac{i (N - 1) \lambda^2}{1 4 
4 \pi} \frac{\sqrt{{\cal M}_\sigma^2 - 4 {\cal M}_\pi^2}}{{\cal 
M}_\sigma} \varphi^2 (X) \nonumber \\ 
&& + \frac{i (N - 1) \lambda^2}{1 4 4 \pi} \frac{\varphi^2 (X)}{p} 
\int_{\xi_{l3}}^{\xi_{u3}} d \xi \, N_\pi (X; \xi, \hat{\bf q}) 
\left[ 2 + N_\pi (X; E_p^{(\sigma)} - \xi, \widehat{{\bf p} - {\bf 
q}}) \right] \, \rule[-3mm]{.14mm}{8.5mm} 
\raisebox{-2.85mm}{\scriptsize{$\; \hat{\bf p} \cdot \hat{\bf q} = 
z_3$}} \nonumber \\ 
\end{eqnarray} 
with 
\begin{eqnarray} 
& \xi_{u1 (l1)} = \frac{{\cal M}_\sigma \left[ {\cal M}_\sigma 
E_p^{(\pi)} \pm p \sqrt{{\cal M}_\sigma^2 - 4 {\cal M}_\pi^2} 
\right]}{2 {\cal M}_\pi^2} \, , \;\;\;\; 
& z_1 = \frac{2 E_p^{(\pi)} \xi - {\cal M}_\sigma^2}{2 p \sqrt{\xi^2 
- {\cal M}_\sigma^2}} \, , \nonumber \\ 
& \xi_{u2 (l2)} = \xi_{u1 (l1)} - E_p^{(\pi)} \;\;\;\; 
& z_2 = \frac{2 E_p^{(\pi)} \xi - {\cal M}_\sigma^2 + 2 {\cal 
M}_\pi^2}{2 p \sqrt{\xi^2 - {\cal M}_\pi^2}} \, , \nonumber \\ 
& \xi_{u3 (l3)} = \frac{{\cal M}_\sigma E_p^{(\sigma)} \pm p 
\sqrt{{\cal M}_\sigma^2 - 4 {\cal M}_\pi^2}}{2 {\cal M}_\sigma} 
\, , \;\;\;\; 
& z_3 = \frac{2 E_p^{(\sigma)} \xi - {\cal M}_\sigma^2}{2 p 
\sqrt{\xi^2 -{\cal M}_\pi^2}}  \, . \nonumber 
\end{eqnarray} 
 
\end{document}